\documentclass[twocolumn]{emulateapj}

\usepackage{color}

\usepackage{natbib}
\slugcomment{\textbf{Accepted for publication in The Astrophysical Journal}}

\shorttitle{Characterization of NLTT 33370}
\shortauthors{J. E. Schlieder et al.}

\begin{document}


\title{Characterization of the Benchmark Binary NLTT 33370\footnotemark[$\dagger$]\footnotemark[*]}


\author{Joshua E. Schlieder\altaffilmark{1,2}, Micka\"el Bonnefoy\altaffilmark{1,3}, T. M. Herbst\altaffilmark{1}, S\'ebastien L\'epine\altaffilmark{4,5}, Edo Berger\altaffilmark{6}, Thomas Henning\altaffilmark{1}, Andrew Skemer\altaffilmark{7}, Ga\"{e}l Chauvin\altaffilmark{3}, Emily Rice\altaffilmark{2,4,8}, Beth Biller\altaffilmark{1,9}, Julien H. V. Girard\altaffilmark{10}, Anne-Marie Lagrange\altaffilmark{3}, Philip Hinz\altaffilmark{7}, Denis Defr\`{e}re\altaffilmark{7}, Carolina Bergfors\altaffilmark{1,11}, Wolfgang Brandner\altaffilmark{1}, Sylvestre Lacour\altaffilmark{12}, Michael Skrutskie\altaffilmark{13}, Jarron Leisenring\altaffilmark{7,14}}

\footnotetext[$\dagger$]{Based on observations collected at the European Organization for Astronomical Research in the Southern Hemisphere, Chile during program ID 090.C-0819}
\footnotetext[*]{The LBT is an international collaboration among institutions in the United States, Italy and Germany. LBT Corporation partners are: The University of Arizona on behalf of the Arizona university system; Istituto Nazionale di Astrofisica, Italy; LBT Beteiligungsgesellschaft, Germany, representing the Max-Planck Society, the Astrophysical Institute Potsdam, and Heidelberg University; The Ohio State University, and The Research Corporation, on behalf of The University of Notre Dame, University of Minnesota and University of Virginia.}
\altaffiltext{1}{Max-Planck-Institut f\"ur Astronomie, K\"onigstuhl 17, 69117, Heidelberg, Germany}
\altaffiltext{2}{Visiting Astronomer, Kitt Peak National Observatory, National Optical Astronomy Observatory, which is operated by the Association of Universities for Research in Astronomy (AURA) under cooperative agreement with the National Science Foundation.}
\altaffiltext{3}{UJF-Grenoble 1 / CNRS-INSU, Institut de Plan\'{e}tologie et d'Astrophysique de Grenoble (IPAG) UMR 5274, Grenoble 38041, France}
\altaffiltext{4}{Department of Astrophysics, Division of Physical Sciences, American Museum of Natural History, Central Park West at 79th Street, New York, NY 10024, USA}
\altaffiltext{5}{Department of Physics and Astronomy, Georgia State University, Atlanta, GA 30302-4106, USA}
\altaffiltext{6}{Harvard-Smithsonian Center for Astrophysics, 60 Garden Street, Cambridge, MA 02138, USA}
\altaffiltext{7}{Steward Observatory, Department of Astronomy, University of Arizona, 933 N. Cherry Ave, Tucson, AZ 85721, USA}
\altaffiltext{8}{Department of Engineering Science and Physics, College of Staten Island, City University of New York, 2800 Victory Blvd, Staten Island, NY 10314}
\altaffiltext{9}{Institute for Astronomy, University of Edinburgh, Blackford Hill View, Edinburgh EH9 3HJ, UK}
\altaffiltext{10}{European Southern Observatory, Casilla 19001, Santiago 19, Chile}
\altaffiltext{11}{Department of Physics and Astronomy, University College London, Gower Street, London, WC1E 6BT, UK}
\altaffiltext{12}{LESIA, Observatoire de Paris, CNRS, University Pierre et Marie Curie Paris 6 and University Denis Diderot Paris 7, 5 place Jules Janssen, 92195 Meudon, France}
\altaffiltext{13}{Department of Astronomy, University of Virginia, Charlottesville, VA 22904, USA}
\altaffiltext{14}{Institute for Astronomy, ETH, Wolfgang-Pauli-Strasse 27, 8093 Zurich, Switzerland}

\email{schlieder@mpia-hd.mpg.de}






\begin{abstract}
We report the confirmation of the binary nature of the nearby, very low-mass system NLTT 33370 with adaptive optics imaging and present resolved near-infrared photometry and integrated light optical and near-infrared spectroscopy to characterize the system.  VLT-NaCo and LBTI-LMIRCam images show significant orbital motion between 2013 February and 2013 April.  Optical spectra reveal weak, gravity sensitive alkali lines and strong lithium 6708~\AA~absorption that indicate the system is younger than field age. VLT-SINFONI near-IR spectra also show weak, gravity sensitive features and spectral morphology that is consistent with other young, very low-mass dwarfs. We combine the constraints from all age diagnostics to estimate a system age of $\sim$30-200 Myr. The 1.2-4.7 $\mu$m spectral energy distribution of the components point toward $\mathrm{T_{eff}=3200 \pm 500}$ K and  $\mathrm{T_{eff}=3100 \pm 500}$ K for NLTT 33370 A and B, respectively. The observed spectra, derived temperatures, and estimated age combine to constrain the component spectral types to the range M6-M8. Evolutionary models predict masses of $113\pm8$ $M_{Jup}$ and $106\pm7$ $M_{Jup}$ from the estimated luminosities of the components.  KPNO-Phoenix spectra allow us to estimate the systemic radial velocity of the binary.  The Galactic kinematics of NLTT 33370AB are broadly consistent with other young stars in the Solar neighborhood. However, definitive membership in a young, kinematic group cannot be assigned at this time and further follow-up observations are necessary to fully constrain the system's kinematics. The proximity, age, and late-spectral type of this binary make it very novel and an ideal target for rapid, complete orbit determination. The system is one of only a few model calibration benchmarks at young ages and very low-masses.  
\end{abstract}





\section{Introduction}
\setcounter{footnote}{0}

Very low-mass (VLM) dwarfs (M $\lesssim$ 0.1 M$_{\odot}$) occupy a complex parameter space of mass, temperature, and atmospheric properties.  In this transition region between stable, hydrogen burning stars and progressively cooling brown-dwarfs, interiors become degenerate and fully convective and atmospheres are dominated by complex molecules and the formation of clouds \citep{Chabrier97, Burrows03}.  Mass is the most fundamental parameter used to determine the properties and evolution of VLM dwarfs, but is undoubtedly also one of the most difficult to obtain.  Observed luminosities and temperatures and reliable ages allow for the inference of mass via comparison to predictions from theoretical models.  However, these models suffer from systematic uncertainties introduced by assumptions made regarding internal and atmospheric physics \citep{Baraffe02, Mathieu2007}.  

Binaries are the key to understanding VLM dwarfs as they provide the rare opportunity to directly determine masses and provide necessary reference points for the direct calibration of models.  VLM binary dynamical masses have been extensively studied via high-resolution adaptive optics (hereafter AO) imaging \citep[][and references therein]{Bouy08, Dupuy10, Kono10} and proved an invaluable resource in understanding the limitations of theoretical models. However, the majority of  VLM dwarf mass determinations are for old, field age targets.  Fewer than ten young ($\lesssim$100 Myr), low-mass ($\le$0.5 M$_{\odot}$) binaries have dynamical mass determinations using imaging or other methods \citep[e.g.][]{Prato02, Close05, Stassun06, Kono07, Bonnefoy09, Schaefer12}.   Model systematic errors are believed to be more severe at the low masses and young ages of these systems because of increased sensitivity to choices of initial conditions and the proper treatment of low-gravity atmospheres \citep{Baraffe02}. Thus, the current sample of calibrators is too limited to provide definitive results regarding the reliability of the models.   

An ongoing astrometric monitoring program of young, low-mass binaries aims to add more benchmarks to this sample \citep[see][]{Bonnefoy09}. One target in this program is NLTT 33370 (2MASS J13142039+1320011).  \citet{Law06} used $z^{\prime}$ and $i^{\prime}$ Lucky Imaging to identify the star as a candidate binary.  Their images showed a very tight, 0\farcs13 separation system with $\Delta$$z^{\prime}$ and $\Delta$$i^{\prime}$$\approx$1.0. The lack of a second observing epoch, and thus verification of common proper motion, prevented confirmation of a bound system. Three years later, \citet{Lep09_2} published a parallax and the first spectrum of the components. They revealed a highly active system at 16.39$\pm$0.75 pc with a combined M7.0$\pm$0.5 optical spectral type.  Radio, spectral, and photometric monitoring were later used to determine that NLTT 33370 is a radio-bright, periodic variable with rapid rotation \citep{McLean11}. \citet{Schlieder12_1} also proposed the system as a kinematic candidate of the $\sim$100 Myr old AB Doradus moving group \citep{Zuck04_1, Baren13}. Thus, NLTT 33370 is an intriguing target that offers the rare opportunity to calibrate models at very low-masses on a short time scale. We have therefore pursued a detailed characterization of the system as part of an ongoing survey to identify and characterize young, low-mass stars in the solar neighborhood (Schlieder et al.~2014, in prep).  We present here a combined analysis of imaging and spectroscopic data.

\section{Observations and Data Reduction}

\subsection{Adaptive Optics Imaging}\label{subsec:nacoim}

\subsubsection{VLT-NaCo}
\label{subsubsection: VLT-NaCO}

We obtained high-resolution $J$-band ($\lambda_{c}$=1.27 $\mu$m) imaging of NLTT 33370 on 2013 February 10 UT with the NaCo instrument \citep{Lenzen03, Rousset03} mounted at the 8.2m VLT UT4. Since the target is faint in the optical, we opted to use the near-IR wavefront sensor to feed the AO system. We acquired 20 $\times$ 5 s exposures of the source with the S13 camera (13.25 mas pixel$^{-1}$ sampling). The source was dithered 6\farcs0 every 6 exposures to subtract sky background and filter out bad pixels. We also observed the star BD+16 2490 with the same setup but different exposure times (0.3454 s) to estimate the point spread function (PSF). For calibration of the platescale and orientation on the chip, we obtained 10 frames of the $\Theta$ Orionis C astrometric field on 09 February 2013 UT \citep{1994AJ....108.1382M}. We repeated the entire $J$-band observing sequence in the $K_{s}$-band ($\lambda_{c}$=2.18 $\mu$m) with the S13 camera on 2013 March 13 UT. We did not however obtain observations of the astrometric field during the same night. Therefore, we analyzed publicly available data of $\Theta$ Orionis C obtained with the same setup as our $K_s$ band data on 16 November 2012 UT\footnote{No changes were made to the instrument since these observations were taken that would have significantly changed the determination of true North.}.  Data were reduced using the ESO $Eclipse$ software \citep{Devillard97} using a method similar to that described in \citet{Bonnefoy09}.  The reduction included standard corrections for bias, sky, and bad pixels and also frame re-centering and averaging.

\subsubsection{LBTI-LMIRCam}

We obtained additional high-angular resolution images of the binary with the near-infrared camera LMIRCam \citep{2008SPIE.7013E.100H, 2010SPIE.7735E.118S, 2012SPIE.8446E..4FL}, a component of the Large Binocular Telescope Interferometer (LBTI), on 19 April 2013 UT. The camera is mounted at the interferometric focus of the LBT and benefits from AO correction provided by the secondary mirrors of the telescope \citep{2010ApOpt..49G.174E, 2011SPIE.8149E...1E}. Only one of the two telescopes was used during the observations\footnote{``Single Aperture Mode", the other second secondary mirror could not be operated at the time of the observations.} and the AO loop was successfully closed on this faint binary \citep[$R$=14.9,][]{Monet2003}.

We obtained 100 $\times$ 4.075 s exposures of the components with the $H$-band filter ($\lambda_{c}=$1.65 $\mu$m, $\Delta \lambda$=0.31 $\mu$m). The target was dithered to two positions in the field of view separated by $\sim$6\farcs5 in order to remove detector bias and allow for background subtraction. We also obtained 118 $\times$ 0.78 s images of the binary with the $L'$-band filter ($\lambda_{c}=$3.70 $\mu$m, $\Delta \lambda$=0.58 $\mu$m). The source was dithered to three positions for these observations.  We observed BD+16 2490 immediately after NLTT 33370AB with the $H$ and $L'$ filters to calibrate the telescope+instrument PSF (900 and $600 \times 0.087$ s exposures, respectively).  We also observed the bright stars HR5384 and $\beta$ Leo in the $L'$-band as additional PSF calibrators.  We lastly obtained 50 $\times$ 0.99 s frames of the M92 astrometric field in the $L'$-band to determine the instrument platescale and orientation. 
  
We followed the procedure described in Section~\ref{subsubsection: VLT-NaCO} to apply cosmetic corrections to the LMIRCam data. The instrument does not have a derotator so individual frames must be re-oriented with true North. The orientation and platescale of the instrument were first determined using images of the wide binary HD 165341 taken at three different orientations on the detector approximately one month after our observations.  The precise astrometry of HD 165341 in the Washington Double Star catalog \citep{2001AJ....122.3466M} allowed us to estimate a platescale of $10.72 \pm 0.01$ mas pixel$^{-1}$ and an orientation of $0.31 \pm 0.20^{\circ}$ East of North.  We confirmed the platescale calibrated with the binary astrometry with our M92 images.  In this case, a reference image of M92 from the Hubble Legacy Archive observed in 2009 with the WFC3 F160W filter\footnote{PropID \#11664, PI T. Brown} was used to calibrate the platescale from measurements of pairwise separations and position angles between 5 stars.   From this analysis, we find an LMIRCam platescale of $10.76 \pm 0.05$ mas pixel$^{-1}$.  We thus adopt a platescale of $10.74 \pm 0.05$ mas pixel$^{-1}$ and a detector orientation of $0.31 \pm 0.20^{\circ}$ East of North for our subsequent analyses.  All of our observed frames were re-oriented with true North using this detector orientation and then re-aligned into a master cube using cross-correlation. Since the LBT AO system has a visible wavefront sensor, the low $R$-band luminosity of the target led to a variable $H$-band Strehl ratio. We therefore selected the best 10\% of the frames in the $H$-band cube and median combined them to produce the final images. We did not apply any frame selection on $L'$-band data where the Strehl was much more stable.

\subsection{Near-Infrared Spectroscopy}
\subsubsection{VLT-SINFONI}

We obtained near-infrared (1.10-2.45 $\mu$m) spectra of NLTT 33370AB with the SINFONI integral field spectrograph \citep{Eisenhauer03,  Bonnet04} on VLT UT4.  Observations were performed on 2013 March 12 UT and 2013 February 26 UT with the $J$ (1.10-1.40 $\mu$m; R$\sim$2360) and $H$+$K$-band (1.45-2.45 $\mu$m; R$\sim$2710) gratings respectively. We used pre-optics that produced 12.5$\times$25 mas rectangular spaxels and a field of view of 0\farcs8 $\times$ 0\farcs8.  The AO system of the instrument (MACAO) was locked and partially resolved the binary. We took 7 and 10 $\times$ 10 s exposures on the target in the $J$ and $H$+$K$-bands respectively. Small dithers (0\farcs2) were used to filter out bad pixels in post-processing. Additional sub-pixel, on-sky dithering allowed for final data cubes with a regular spatial sampling of 12.5 mas pixel$^{-1}$.  At the end of each sequence, the telescope was moved to an empty field in order to estimate the sky background. 

Each raw frame was first corrected for noise created by the detector electronics using recipes described in the ESO data reduction cookbook version 1.0\footnote{http://www.eso.org/sci/facilities/paranal/instruments/               sinfoni/doc/VLT-MAN-ESO-14700-4037.pdf}. Datacubes were then built from each corrected raw frame of the binary using the 2.3.2 release of the ESO reduction pipeline \citep{Abuter06}.  A detailed description of the pipeline procedures and further post-processing is available in \citet{2013arXiv1306.3709B}. Here, we summarize the results.  The output of the pipeline is a merged data cube with a 1\farcs225 field of view. Spectra were extracted by integrating the flux in each cube plane over circular apertures. Telluric corrections were applied using B-type standards observed during the same run at similar airmass. The final $J$ and $H$+$K$-band spectra of NLTT 33370AB were flux-calibrated and combined into a single spectrum spanning 1.1-2.45 $\mu$m using the binary's 2MASS photometry \citep{Cutri03} and the 2MASS flux zero points \citep{Cohen03}.

\subsubsection{KPNO Phoenix}

\citet{Schlieder12_1} suggested that NLTT 33370AB has proper motion and distance consistent with other proposed members of the AB Doradus moving group. This is only an indication of possible membership to the group and to fully investigate the binary's 6D Galactic kinematics (UVWXYZ), we sought to measure the radial velocity (RV) of the system.   We obtained two spectra of NLTT 33370AB on 2013 January 04 and 05 UT with the Phoenix near-IR spectrograph \citep{Hinkle98} on the KPNO 4.0m Mayall telescope.  

We observed in the $H$-band, centered at 1.56 $\mu$m, following the observing procedure described in \citet{Schlieder12_3}.  Integration times of 800 and 630 s were used for the two observations respectively.  We reduced the data using standard IRAF\footnote{IRAF is distributed by the National Optical Astronomy Observatory, which is operated by the Association of Universities for Research in Astronomy (AURA) under cooperative agreement with the National Science Foundation.} routines, which included background subtraction, flat-fielding, bad pixel removal,  and wavelength calibration using OH night sky lines.  The individual spectra from each night were then averaged to boost the median signal-to-noise ratio (SNR) to $\sim$20.  

\section{Analyses}
\subsection{Imaging Data}
\subsubsection{VLT-NaCo Analysis}

\begin{figure*}[!h]
\epsscale{1.0}
\plotone{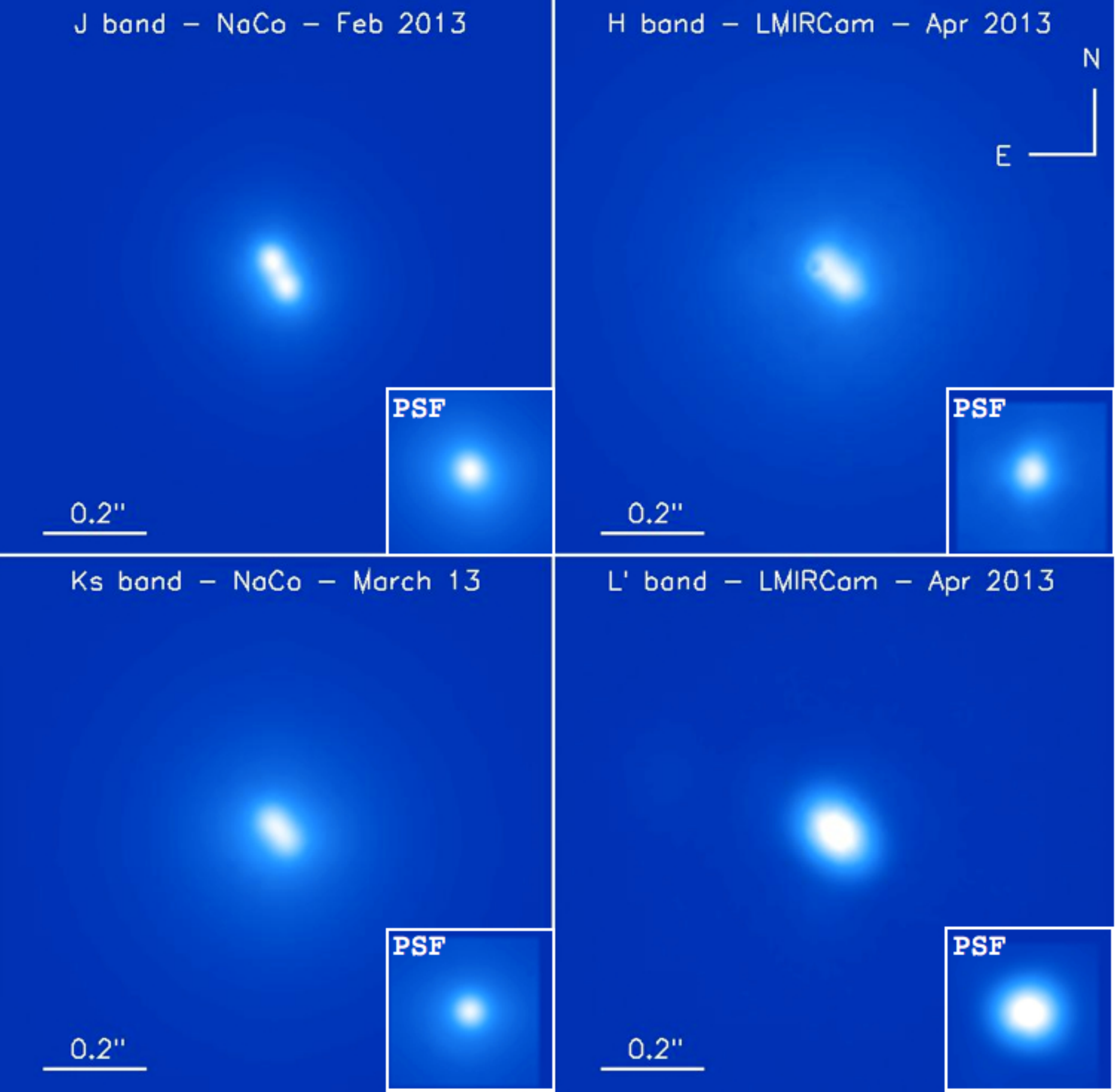}
\caption{VLT-NaCo $J$ (top-left) and $K_{s}$ (bottom-left) and LBT-LMIRCam $H$ (top-right) and $L'$-band (bottom-right) images of NLTT 33370 A and B. The separation of the components is $\rho=0\farcs076\pm0\farcs005$ in the $J$-band images. The different epochs, separated by only two months, reveal the rapid orbital motion of the binary. \label{figimage}}
\end{figure*}

The binary is clearly re-resolved in the $J$-band images and is blended in the $K_{s}$-band due to poorer conditions at the time of those observations (Figure~\ref{figimage}, left column). These follow-up images verify the binary nature of the system.  The proper motion of NLTT 33370AB is so large that in the $\sim$8 years since the discovery image, a background object would have moved about 2$^{\prime\prime}$.    Component fluxes were deconvolved and positions determined using the algorithm proposed by \citet{Veran98} and images of BD+16 2490 as input PSFs.  We find the components have essentially equal flux in the $J$ and $K_s$-bands ($\Delta J = 0.10 \pm 0.11$ mag, $\Delta K_s = 0.10 \pm 0.21$ mag).  We combined the 2MASS magnitudes of the system with these contrasts to estimate the J and $K_s$-band photometry of the binary components (see Table~\ref{tab1}).  The equal flux of the components is indication that they are nearly equal spectral type/mass.  We also derive the position angle and separation of the binary in each band, corrected for the instrument orientation and platescale.  The components are separated by $\rho=76\pm5$ mas in both the $J$ and $K_s$-band images (projected separation 1.24$\pm$0.11 AU).   

Our indistinguishable component magnitudes make derivation of the true position angle (PA) difficult.  In our $J$-band images, we measure PA $=204.9\pm0.3^{\circ}$ and in our $K_s$-band images we measure PA $=215.2\pm1.0^{\circ}$. The PA values in each band reported here, and in later sections of this manuscript, may be 180$^{\circ}$ of out phase due to our uncertainty in assigning the components.  We discuss the PA of the system further in the next subsection.  Regardless of the true PA, we measure $\sim$10$^{\circ}$ of orbital motion in only one month; revealing rapid orbital motion.

\subsubsection{LBTI-LMIRCam Analysis}

With these LBT observations, we aimed to better understand the orbital motion observed in the NaCo data and provide more reference points in the IR spectral energy distributions (SEDs) of the components.  The binary was marginally resolved in the $H$-band and exhibits an elongated PSF in the $L'$-band images (Figure \ref{figimage}, right column). The blending of the components was deconvolved (see section \ref{subsubsection: VLT-NaCO}) using primarily the images of BD+16 2490. We also used images of HR5384 and $\beta$ Leo as alternative PSFs in the $L'$-band to evaluate the impact of the PSF on the deconvolution process. 

\begin{figure}[h]
\epsscale{1.0}
\plotone{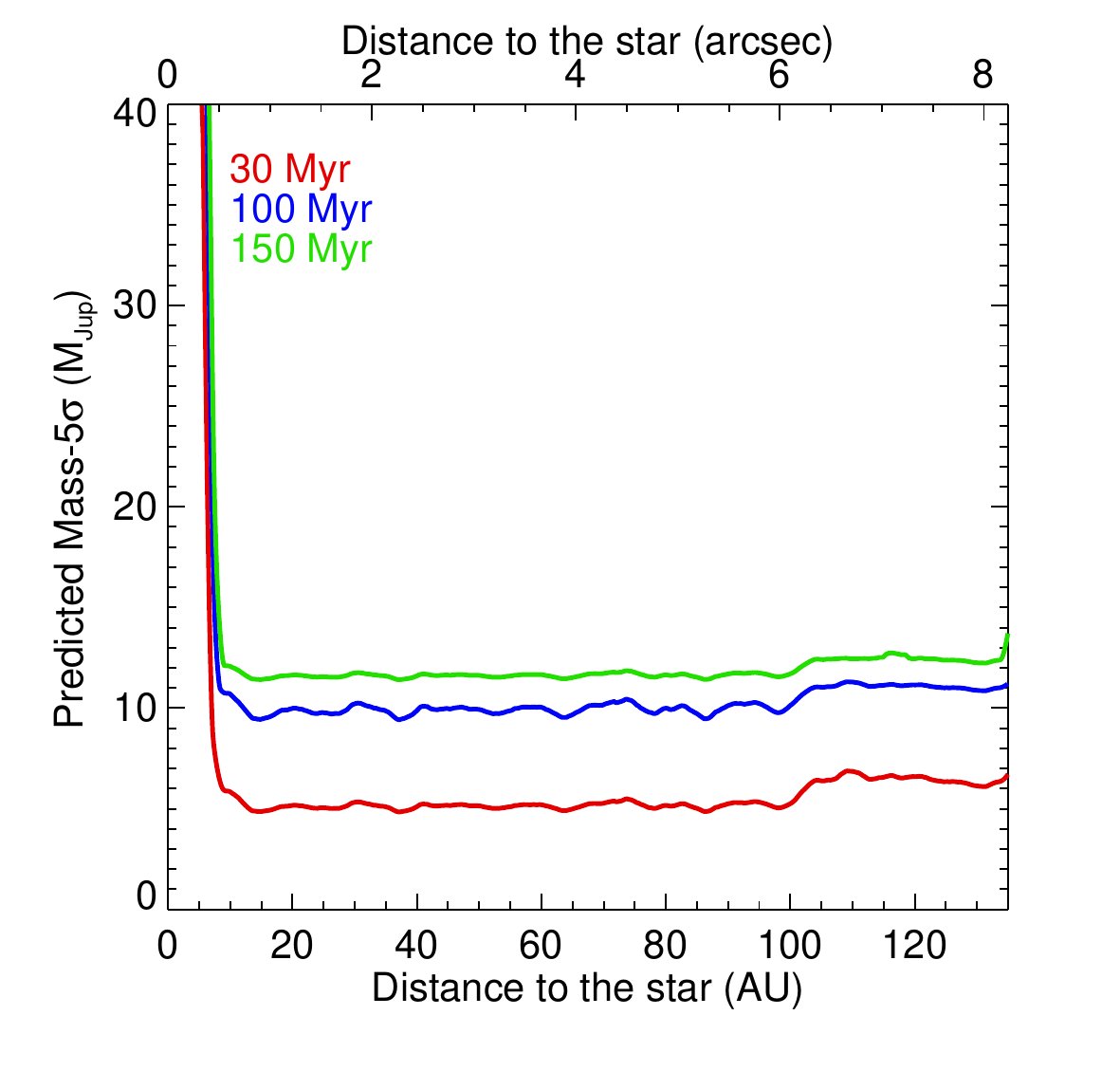}
\caption{LBTI-LMIRCam $L'$-band detection limits for NLTT 33370AB converted to masses using the COND03 evolutionary tracks \citep{Baraffe03}. \label{figlcon}}
\end{figure}

The deconvolution yields contrasts $\Delta H = 0.03 \pm 0.06$ mag and $\Delta L' = 0.03 \pm 0.15$ mag. The components are again indistinguishable. We used the 2MASS $H$-band magnitude of the unresolved pair to estimate the magnitude of each component. In the $L'$-band, WISE W1 photometry \citep{Wright2010, Cutri12} was used as an estimate of the system magnitude to retrieve individual component magnitudes. We report the final individual photometry in Table \ref{tab1}. We used the estimated detector orientation and platescale of LMIRCam to measure PA $=226.7\pm1.6^{\circ}$ and  $\rho=73 \pm 4$ mas in the $L'$-band images. We also find PA $=224\pm1^{\circ}$ and $\rho=65\pm9$ mas in our $H$-band data. However, we find $H$-band astrometry less reliable because of the frame selection used to construct the final image and do not recommend their use for orbit analyses. We do not detect additional companions at larger separations in the $L'$-band observations. We report the detection limit, converted to companion masses for three possible system ages (see Section~\ref{age}) in Figure~\ref{figlcon}. 

The measured separations and position angles from the LMIRCam data again reveal $\sim$10$^{\circ}$ of orbital motion over one month.  Additionally, we measure no significant change in separation for the observed changes in PA.  This result, combined with previous astrometric measurements in \citet{Law06} can place constraints on our PA measurements and provide preliminary information about the orbit of the system.  Our observation of little or no change in separation for $\sim$20$^{\circ}$ of PA change over 2 months indicates an edge-on view of the system is unlikely. Also, \citet{Law06} report a June 2005 epoch separation $\rho$ = 130 $\pm$ 20 mas and PA = 46.0 $\pm$ 2.0$^{\circ}$. This PA estimate is reliable since their data do not suffer the 180$^{\circ}$ PA ambiguity ours do; the components have larger contrasts at visible wavelengths and are distinguishable in their images.  Since our measured separation is $\sim$half that of the earlier epoch and we measure parameters suggestive of a low inclination, we find it plausible that we have observed the system approximately 180$^{\circ}$ out of phase from the discovery epoch, but cannot completely rule out the opposite scenario.  Thus, we favor the larger of possible position angles (i.e.~around 200$^{\circ}$) and report them in Table~\ref{Table:astrometry}.  These combined results are also preliminary indication that the orbit may be eccentric.  In our subsequent analyses, we refer to components NLTT 33370A and NLTT 33370B, but the derived parameters all overlap within uncertainties and we refrain from explicitly assigning a primary component in our images.

The projected separation in the 2005 epoch (2.16 AU) implies an orbital period of $\sim$6.8 years if we use the component masses estimated in Section~\ref{phys_param} (see Table~\ref{tab1}). This is consistent with the rapid orbital motion of $\sim$10$^{\circ}$ per month derived from our imaging observations. When compared with the astrometry of the discovery image, our measurements indicate that the binary could be halfway through its second orbit since 2005. Nevertheless, additional astrometric epochs are needed to understand the full orbital parameters of the system.

\subsection{Spectroscopic Data}
\subsubsection{Previously Published Optical Spectra}

\begin{figure*}[!t]
\epsscale{1.0}
\plotone{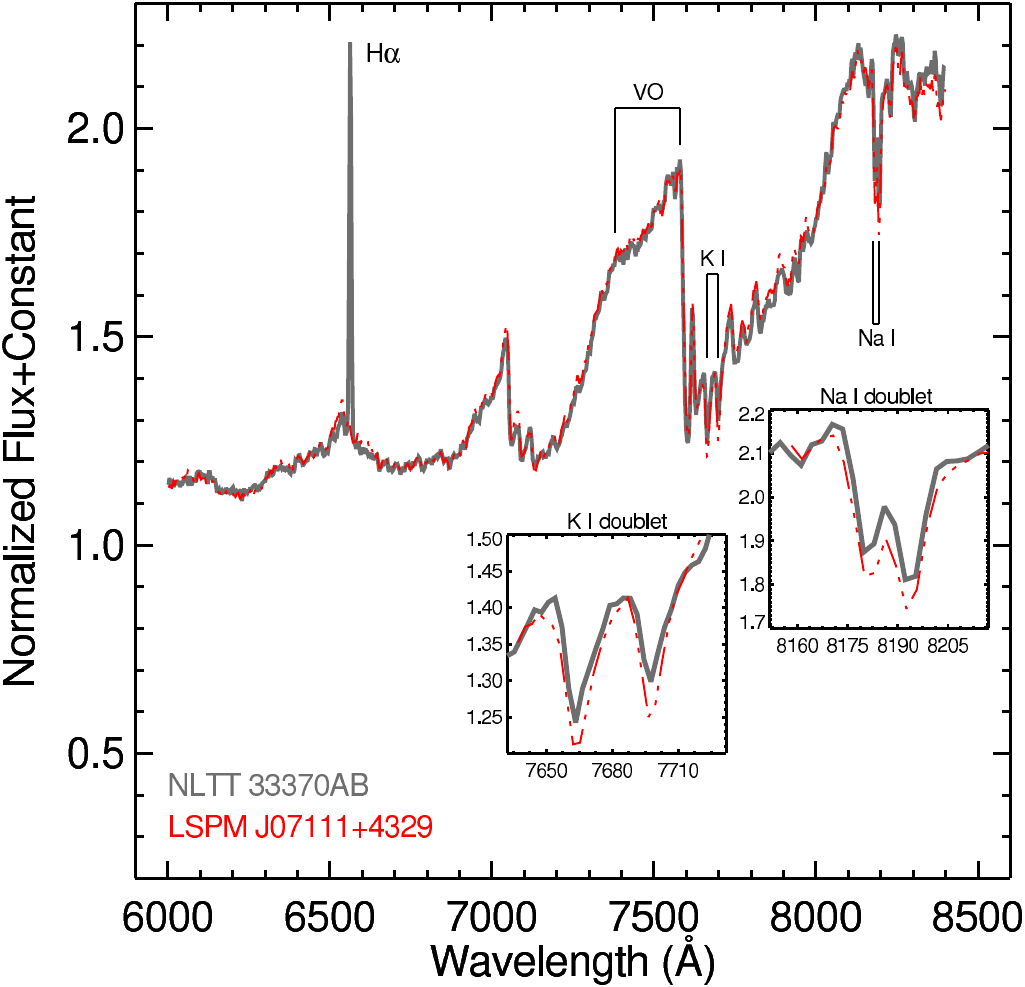}
\caption{Comparison of medium resolution optical spectra of NLTT 33370AB (gray, solid) and LSPM J07111+4329 (red, dashed). The spectra are nearly identical except for features indicative of activity and low surface gravity.  The insets show two alkali doublets that are sensitive to surface gravity.   These lines are visibly weaker in NLTT 33370AB than in the comparison dwarf. The 4.8$\pm$0.8~\AA~equivalent width of the Na I doublet is consistent with an age younger than the field but older than Upper Scorpius. \label{figolow}}
\end{figure*}

We re-analyzed the medium-resolution optical spectrum of NLTT 33370AB, initially described in \citet{Lep09_2}.  The spectrum was obtained on UT 2006 January 30 UT with the MkIII spectrograph and ``Wilbur'' CCD camera on the 1.3m McGraw-Hill telescope at the MDM observatory.  We performed a visual comparison to a field age, M7V binary observed with an identical setup when the instrument was on the MDM 2.4m Hiltner telescope in 2003.  We aimed to constrain the age of NLTT 33370AB by investigating possible differences in alkali lines and molecular bands affected by surface gravity \citep[e.g.][]{McGovern04, Allers07, Kirkpatrick08, Cruz09, Rice10}.  Fig.~\ref{figolow} shows the comparison, where both spectra are normalized to the continuum at 8100~\AA.  The overall match is striking: near perfect coincidence across the continuum and broad TiO bands, with an obvious deviation around the -54.1~\AA~equivalent width (EW, negative width convention for emission) H$\alpha$ feature (6563~\AA) of NLTT 33370AB \citep{Lep09_2}.  

The insets in Fig.~\ref{figolow} show the regions around the neutral potassium (K I) 7665~\AA~and 7699~\AA~doublet and the neutral sodium (Na I) 8183~\AA~and 8195~\AA~doublet.  Both of these features are alkali lines whose widths are influenced by photospheric gravity, in the sense that lower surface gravity (for dwarfs, a younger age) leads to weaker lines. Compared to the older binary with identical type, NLTT 33370AB has visibly weaker K I and Na I doublets.  We measured a 4.8$\pm$0.8~\AA~EW for the Na I doublet using the method and integration regions described in \citet{Schlieder12_2}. This EW is more than 1~\AA~smaller than that of the the comparison spectrum and consistent with an age younger than the field \citep[see][their Fig.~3]{Schlieder12_2}. The width of the Na I doublet is also consistent with an intermediate age between the field and Upper Scorpius \citep[U Sco, 11$\pm$2 Myr][]{Pecaut12} when placed in the empirical sample of VLM dwarfs presented in \citet{Martin10}. We do not measure the EW of the K I doublet due to the difficulty in selecting a pseudo continuum.  The VO band around 7500~\AA~is also gravity sensitive in the opposite sense of the alkali lines (deeper band, lower gravity).  The VO band of NLTT 33370AB appears marginally stronger than that of the comparison dwarf; again suggestive of lower surface gravity.  The H$\alpha$ emission is not useful as an age indicator in M dwarfs of this late-type, since they remain active for billions of years \citep{West11}.

\begin{figure}[!htb]
\epsscale{1.0}
\plotone{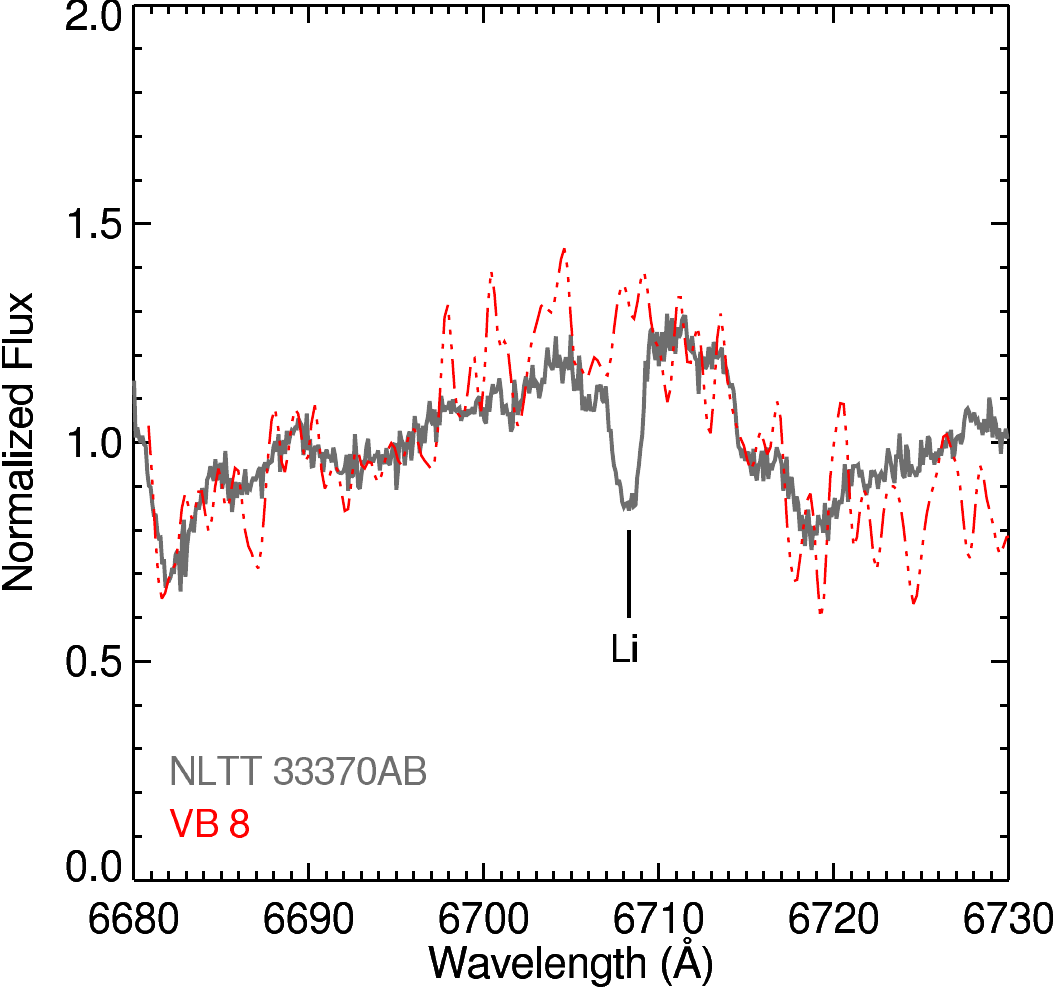}
\caption{The region around the Li 6708~\AA~line in NLTT 33370AB (gray, solid) compared to the field M7 VB 8 (red, dashed). Li absorption is evident in our target with an equivalent width (EW) of $\sim$500 m\AA.  This EW is comparable to those observed in other young, VLM dwarfs in the Pleiades and young, moving groups and is evidence that the system is too young, or at least one of the components is not massive enough, to have depleted its primordial Li abundance. \label{figlith}}
\end{figure}

We also performed an analysis of the high-resolution optical spectrum of NLTT 33370AB first reported in \citet{McLean11}.  The spectrum was obtained on 2009 May 19 UT using the Magellan Inamori Kyocera Echelle (MIKE) spectrograph on the Magellan Clay 6.5m telescope. The evidence for youth in the medium resolution spectrum prompted us to search at high-resolution for lithium absorption at 6708~\AA~as another probe of a young age.  Figure~\ref{figlith} shows a portion of the continuum normalized spectrum around the Li 6708~\AA~line in NLTT 33370AB  compared to a spectrum of the field age M7 dwarf, VB 8 \citep[from][]{Tinney98}. Lithium is clearly seen in absorption in NLTT 33370AB compared to the field standard. We measured a Li EW of $\sim$460 m\AA\footnote{Li EW not corrected for possible contamination from the nearby Fe I line at 6707.4~\AA}. This is comparable to Li EW's measured for VLM Pleiades and IC 2391 members with similar spectral type \citep{Stauffer98, Barrado04} and for the M6 binary TWA 22AB in the $\sim$10-20 Myr old $\beta$ Pic moving group \citep{Mentuch08}.  2MASS J05575096-1359503, a $\sim$10 Myr old M7 dwarf studied in \citet{Shkolnik09}, also exhibits a similar Li EW.  The detection of Li in the spectrum of NLTT 33370AB indicates that the system is either too young to have burned its primordial abundance or at least one of the components does not have sufficient mass to burn Li.  We discuss further the implications of the Li detection in Section~\ref{phys_param}. 

We also note that \citet{McLean11} report an H$\alpha$ EW of -9.9~\AA~from their high-resolution spectrum. This value is 5$\times$ smaller than the EW of the line shown in Fig.~\ref{figolow}.  To probe variability in the H$\alpha$ line, \citet{McLean11} observed NLTT 33370AB six times during the night of 2009 February 25 UT with the Low Dispersion Survey Spectrograph on the Magellan Clay 6.5m telescope.  They found no variability over time scales of a few hours and an average EW of -14.6~\AA. Thus the H$\alpha$ emission in the spectrum shown in Figure~\ref{figolow} is not representative of the average emission, and in this case, the binary was likely observed during a heightened state of activity.     

\subsubsection{VLT-SINFONI Analysis}
\label{SINFONIanalysis}

\begin{figure}[!h]
\epsscale{0.9}
\plotone{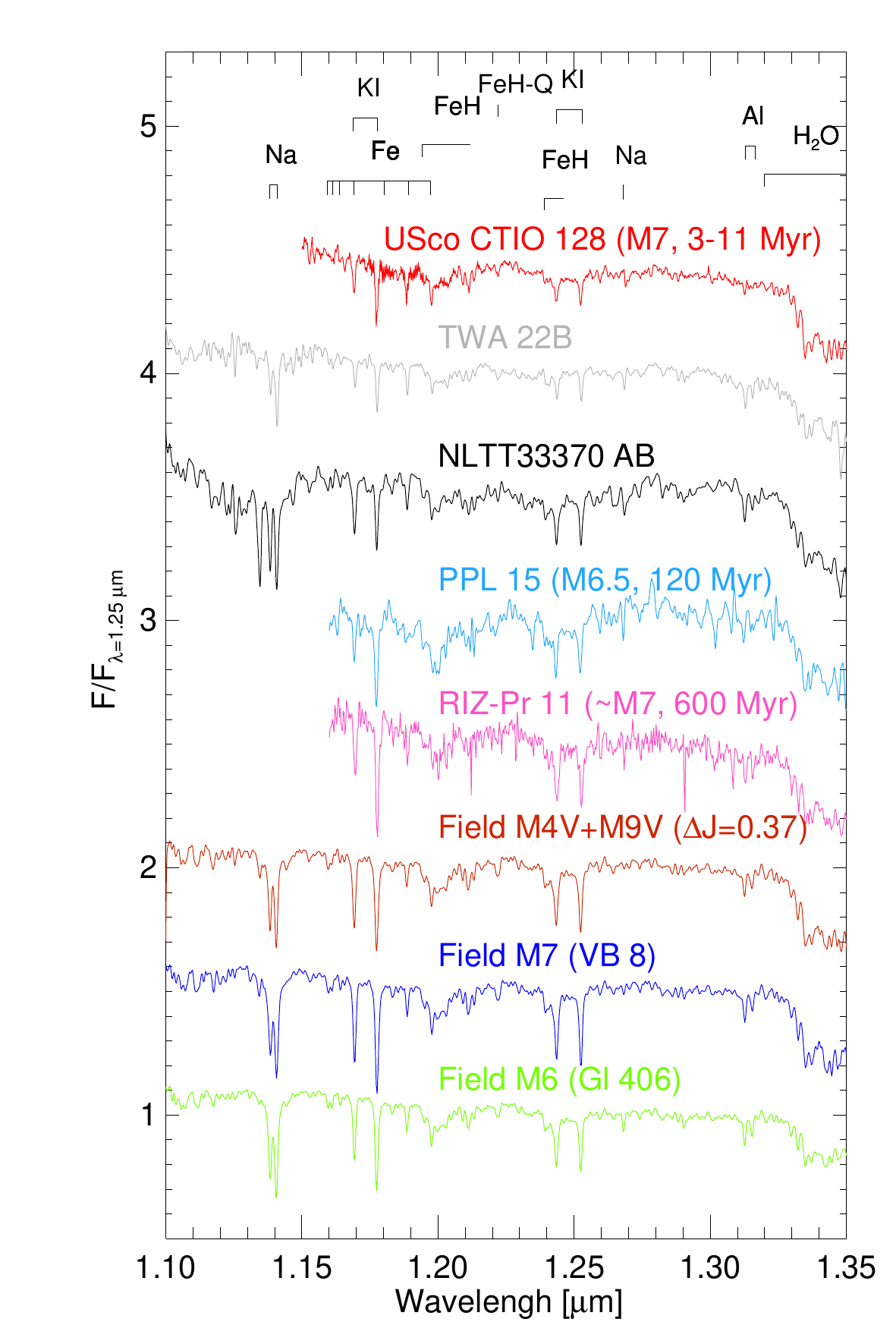}
\caption{Comparison of the SINFONI $J$-band spectrum of NLTT 33370AB to those of field, intermediate, and young age dwarfs. All spectra have been degraded to R$\sim$1400 and normalized at 1.25 $\mu$m.  The spectral slope and H$_{2}$O band long ward of 1.33 $\mu$m are consistent with the other M7 templates.  The strength of the water band is not compatible with earlier spectral types. The the K I doublets and FeH band (1.2 $\mu$m) described in the text are intermediate in strength between the Upper Scorpius and the Praesepe templates.  The spectrum is most similar to PPL 15, a VLM spectroscopic binary in the Pleiades ($\sim$130 Myr). The maroon spectrum is a weighted combination of M4V and M9V field age templates.  This combination is a best least-squares fit to our NLTT 33370AB spectrum. However, the strengths of some features and the $J$-band contrast ratio are inconsistent with our observations.  We therefore conclude that the peculiar features in the $J$-band spectrum of NLTT 33370AB are indicative of reduced surface gravity and a young age.} \label{figjspec}
\end{figure}

\begin{figure}[!h]
\epsscale{0.9}
\plotone{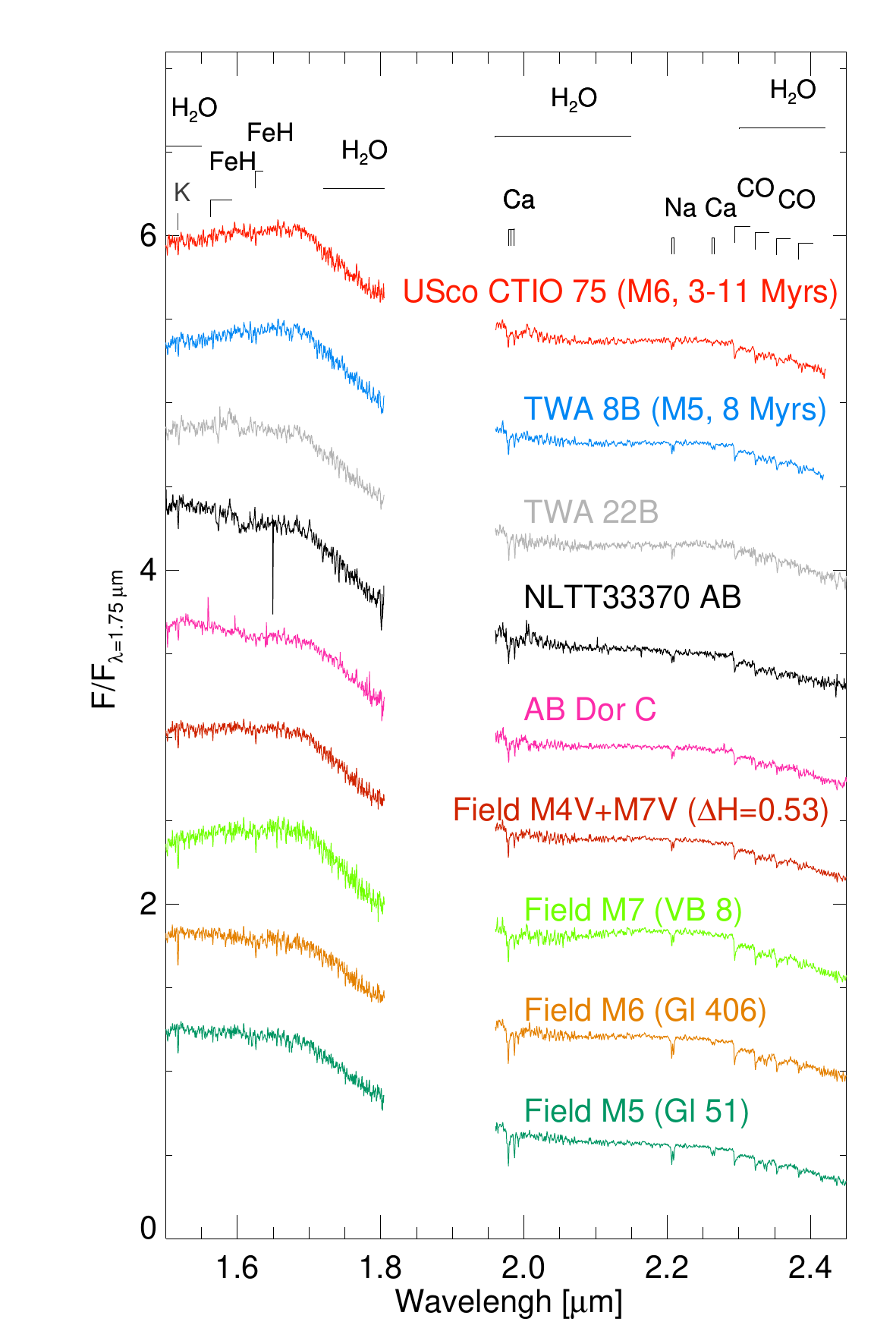}
\caption{$H$+$K$-band spectral sequence for young to field age VLM dwarfs. The SINFONI $H$+$K$-band spectra of NLTT 33370AB are shown in black.  The $H$-band continuum shape is less peaked than the $\sim$10 Myr members of Upper Scorpius (U Sco) and the TW Hydrae association (TWA), suggesting the binary is older.  The overall spectral morphology of NLTT 33370AB is comparable to AB Dor C, an M5 member of the $\sim$100 Myr AB Doradus moving group. Gravity sensitive alkali lines described in the text are also visibly weaker in NLTT 33370AB than in the field age dwarfs.  The maroon spectrum is a weighted combination of M4V and M7V field age templates.  This combination is a best least-squares fit to our NLTT 33370AB spectrum. However the $H$-band continuum shape and the contrast ratio are inconsistent with our observations. Thus, the peculiar features of NLTT 33370AB are again consistent with a young to intermediate age. \label{fighkspec}}
\end{figure}

We compare the spectrum of NLTT 33370AB to those of reference late-M dwarfs from the IRTF spectral library \citep{Cushing05, Rayner09} and M6 and M7 dwarfs from young clusters and young nearby associations \citep{Bonnefoy09, Close07, Slesnick04, Allers10} in Figures~\ref{figjspec} and~\ref{fighkspec}. A visual comparison of the $J$-band spectrum to other late-M dwarf standards reinforces the M7 optical spectral type determination. The spectral slope and strength of the H$_2$O band longward of 1.33 $\mu$m are most consistent with the M6.5-M7 templates. These features are inconsistent with the M6 comparison spectra.  Further, comparison among the M6.5-M7 spectral types reveals that NLTT 33370AB exhibits weak, gravity sensitive K I doublets (1.169 $\mu$m and 1.177 $\mu$m and 1.243 $\mu$m and 1.253 $\mu$m) and a weak FeH band at 1.2 $\mu$m.  These features are stronger than those of the young, M7 U Sco ($\sim$10 Myrs) member, slightly weaker than the Praesepe reference dwarf \citep[$590^{+150}_{-120}$ Myrs,][]{Fossati08}, and most consistent with the Pleiades member PPL 15 \citep[130$\pm$20 Myrs, also an unresolved binary,][see Fig.~\ref{figjspec}]{Barrado04,1999AJ....118.2460B}\footnote{Spectrum of PPL 15 from the Keck Observatory Archive, PID: H222N2L, PI: M. Liu}.  This suggests that the binary has an intermediate age comparable to the Pleiades.

In Figure~\ref{fighkspec}, the shape of the $H$-band continuum is not obviously ``triangular" like the $\sim$10 Myr templates, but it is also not as flat as in the field age templates.  The gravity sensitive Na I doublet at 2.218 $\mu$m also has an intermediate strength compared to M5-M7 field dwarfs and young M6 dwarfs \citep{Allers10}.   The $H$+$K$ band spectrum of the binary is in fact most similar to the SINFONI spectrum of AB DorC (M5.5) described by \citet{2007MNRAS.378.1229T} and \citet{Close07}.  Thus, our visual comparisons of the combined $J$ and $H+K$ spectra to templates at different ages indicate that NLTT 33370AB is older than $\sim$20 Myr, clearly younger than 600 Myr, and most similar to dwarfs with ages $\sim$100 Myrs. 

\citet{Kirkpatrick08} place a more stringent constraint on the upper age limit of VLM dwarfs from gravity sensitive features.  They note that early-L type members of the Pleiades exhibit peculiar spectral features, but members of the Ursa Majoris group \citep[300-600 Myr,][]{Soderblom93, King03} do not. They conclude that $\gtrsim$100 Myrs is the age where peculiar features due to reduced surface gravity start to fade.  Although we lack late-M comparison spectra with ages between the Pleiades and Praesepe (a range of $\sim$400 Myrs), Figs. \ref{figjspec} and \ref{fighkspec} show that the evolution of spectral features with surface gravity is comparable in late M-dwarfs. Thus, our visual inspection of the near-IR spectrum of NLTT 33370AB combined with previous observational results suggests a young to intermediate system age between $\sim$30-300 Myr.

Our use of an unresolved near-IR spectrum may have unforeseen effects on our age interpretation from gravity sensitive features.  As an additional check of our interpretation, we considered the possibility that features indicative of youth in the unresolved SINFONI spectrum of NLTT 33370AB could be created by the mixture of two field age spectra at different temperatures. We simulated unresolved near-IR spectra of the binary by taking the weighted mean of two medium resolution (R$\sim$2000) spectra from a sample of M4V, M4.5V, M5V, M6V, M6.5V, M7V, M8V, and M9V dwarfs in the IRTF spectral library \citep{Cushing05, Rayner09}.  We computed weights corresponding to contrasts $\Delta$J$=0.10\pm0.50$ mag or $\Delta$H$=0.03\pm0.50$ mag, with 0.02 mag increments. The models were normalized to the NLTT 33370AB spectrum and compared in the $J$ (1.10-1.35 $\mu$m), $H$ (1.45-1.80 $\mu$m), $K_s$ (1.96-2.45 $\mu$m), $H+K_s$ (1.45-2.45), and $J+H+K_s$ (1.10-2.45 $\mu$m) bands using a least-squares fitting routine.

A template made from a combination of M4V and M9V dwarf spectra with $\Delta$J$=0.37$ mag reproduce the slope and H$_2$O band of our unresolved $J$-band SINFONI spectrum.  The K I doublets are not well reproduced, they remain weaker in the spectrum of NLTT 33370AB (see Figure~\ref{figjspec}).  Additionally, the $J$-band contrast ratio from the best-fit is incompatible with that derived from our resolved photometry. In the $H$ and $K_s$-bands, a mixture of M4V+M7V template spectra with $\Delta$H$=0.53$ mag also provide a good fit of the pseudo-continuum of NLTT 33370AB. However, the Ca I, Mg I,  and Na I lines appear stronger in the template. The template also produces a redder slope than the binary spectrum from 1.45 to 1.65 $\mu$m (Figure~\ref{fighkspec}), and 1.10-2.45 $\mu$m. Thus, we conclude that our simulations do not succeed in consistently and simultaneously reproducing the unresolved spectrum of the binary across the $JHK_s$-bands. The best fit template mixtures are also not able to reproduce the observed contrast ratios in the near-IR.  We therefore find it unlikely that the observed peculiar features are produced only by an unresolved, field age binary and can be attributed to the low surface gravity of the components.

\begin{figure}[!h]
\epsscale{0.7}
\plotone{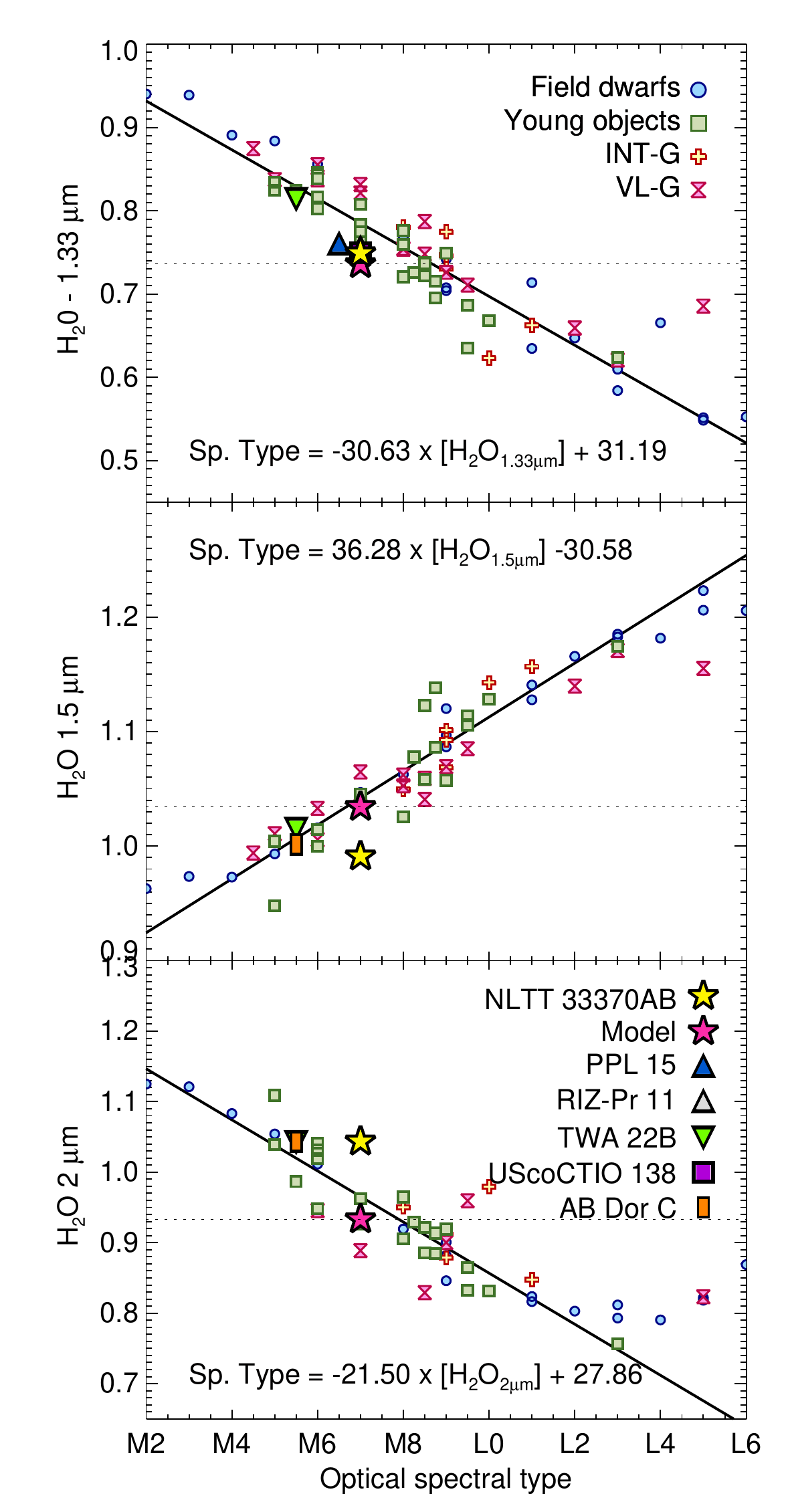}
\caption{ Near-IR H$_{2}$O indices for young to field age VLM dwarfs.  The calculated indices for NLTT 33370AB are shown as a yellow star.  Other symbol designations are given in the figure legend and the samples are detailed in the text. The magenta star labeled ``Model" represents the simulated binary shown in Figure~\ref{figjspec} and Figure~\ref{fighkspec}. The plus signs and hour glasses represent VLM dwarfs with very-low gravity (VL-G) and intermediate gravity (INT-G) classifications from \citet{Allers2013}.  The indices yield spectral types that are generally consistent with optical classifications. The equations provide the linear fits to the data.  Using these equations, we calculate an average near-IR spectral type of M6.5 for NLTT 33370AB, consistent with its optical classification.}  \label{figSpTy}
\end{figure}

To check our visual spectral type classification, we use the H$_{2}$O indices used by \citet{2013arXiv1306.3709B}.  These indices use water absorption features that are weakly sensitive to changes in surface gravity to quantitatively estimate spectral types. The indices we use are very similar to those originally introduced by \citet{Allers07} and used by \citet{Allers2013}.  However, we tuned them to avoid regions of the spectrum typically affected by telluric residuals.  We show the results of this analysis in Figure~\ref{figSpTy}.  We compare the optical spectral types of field dwarfs \citep{Cushing05}, young dwarfs \citep{2013arXiv1306.3709B, Allers2009, Allers07, 2008MNRAS.383.1385L, Slesnick04, Gorlova2003} and standards to the H$_{2}$O indices.  We also show the same comparison for dwarfs with optical spectral types reported in \citet{Allers2013} for which we had near-IR, medium resolution (R$>$700) spectra.  All spectra were degraded to a common resolution of R=300 and interpolated to a common wavelength scale before calculating the indices.  The figure shows that the optical spectral types are generally in good agreement with the trends in the indices.  There is significant scatter from the linear fit, but this scatter is consistent over the plotted spectral type range.  The H$_{2}$O indices of NLTT 33370AB and the other comparison dwarfs are generally in good agreement with their optical types. For NLTT 33370AB, the index around 1.33 $\mu$m yields a later type than the optical classification and the indices at 1.5 $\mu$m and 2.0 $\mu$m yield earlier spectral types.  The average spectral type of NLTT 3370AB from these three indices is M6.5, consistent with the optical classification and our inference from direct comparison to standards in Fig.~\ref{figjspec}.

\subsubsection{KPNO-Phoenix Analysis}

We measure radial velocities via cross-correlation with template spectra from \citet{Prato02} using software written in IDL (Chad Bender, private communication). We first verified the reliability of our RV measurements and checked for possible instrumental errors by measuring the RVs of HD 10870 and HD 73667, two early-K type standards \citep{Chubak12} observed during the same nights as NLTT 33370AB.  The standards' average RVs matched the literature values within uncertainties when cross-correlated against 4 templates with similar spectral type.  The uncertainty in our RV measurement is a combination of the error in a Gaussian fit to the peak of the correlation function and a systematic error of 1 km s$^{-1}$ introduced by the use of empirical templates \citep{Prato02}.    

We first attempted to measure the RV of NLTT 33370AB by correlating against GL 644C, an M7 template.  The correlation function was very complex, with multiple structures of approximately the same correlation power. However, the strongest feature did exhibit two small peaks; preliminary indication of an unresolved binary.   Attempts to perform 2D cross-correlations with different combinations of M6-M9 templates to measure individual component velocities led to unphysical results. The SNR of the input spectrum was simply too low.  We therefore used a Gaussian kernel to bin the spectrum by 3, 5, and 7 pixels. This led to an increase in SNR but a reduction in spectral resolution.  To test for systematic effects introduced by this procedure, we performed it on our RV standard spectra.  The RVs measured from the binned standard spectra were identical to the those of the unbinned spectra within our combined measurement and systematic uncertainties. We found that the NLTT 33370AB spectrum smoothed by 7 pixels produced the strongest correlation features with a consistently strong feature exhibiting a blue shifted, asymmetric peak at $\sim$-19 km s$^{-1}$ or two peaks at $\sim$-19 and $\sim$1 km s$^{-1}$, regardless of the M5-M9 templates used.

Two-dimensional cross-correlations of the binned spectrum with pairs of M6.5/M7 templates (LHS 292, GL 644C, LHS 2351) allowed us to resolve sharp, central peaks in the primary and secondary correlation functions. We measured average RVs of -20.2$\pm$1.2 km s$^{-1}$ and -0.5$\pm$1.5 km s$^{-1}$ respectively.  In the case of these approximately equal mass components, the systemic velocity is the average of their respective RV measurements; $\mathrm{RV_{sys} \approx -10.4\pm1.0}$ km s$^{-1}$. 

If the system RV is close to -10 km s$^{-1}$, the resultant RV amplitude of the components is broadly consistent with the 6.1$\pm$0.5 km s$^{-1}$ amplitude expected for the projected separation measured from our AO imaging and the component masses estimated in Section~\ref{phys_param}.  An eccentric orbit passing close to periastron at the time of the observations may be able to explain both the observed orbital motion and large RV variation between the components.  We caution that our RV measurements are preliminary, and from a single epoch of a binary that exhibits rapid motion in an orbit where the full parameters are unknown.  Future follow-up and measurements from spectra with higher SNRs will allow these kinematic parameters to be much better constrained.  We discuss further the implications of NLTT 33370AB's estimated RV in Section~\ref{kinematics}. 

\section{Discussion}
\subsection{System Age}\label{age}

Each of our spectral analyses suggest that NLTT 33370AB is younger than field age. Our comparison of gravity sensitive features in optical and near-IR spectra to standards at different ages constrain the age to $\sim$30-300 Myr.  In addition, the $J$-band and $H$+$K$-band spectra in Fig.~\ref{figjspec} and Fig.~\ref{fighkspec} are most similar to PPL 15 and AB DorC, which is suggestive of a comparable $\sim$100 Myr age. In the following subsections we present detailed analyses of multiple age indicators.  

\subsubsection{Surface Gravity}

\begin{figure}[!h]
\epsscale{0.7}
\plotone{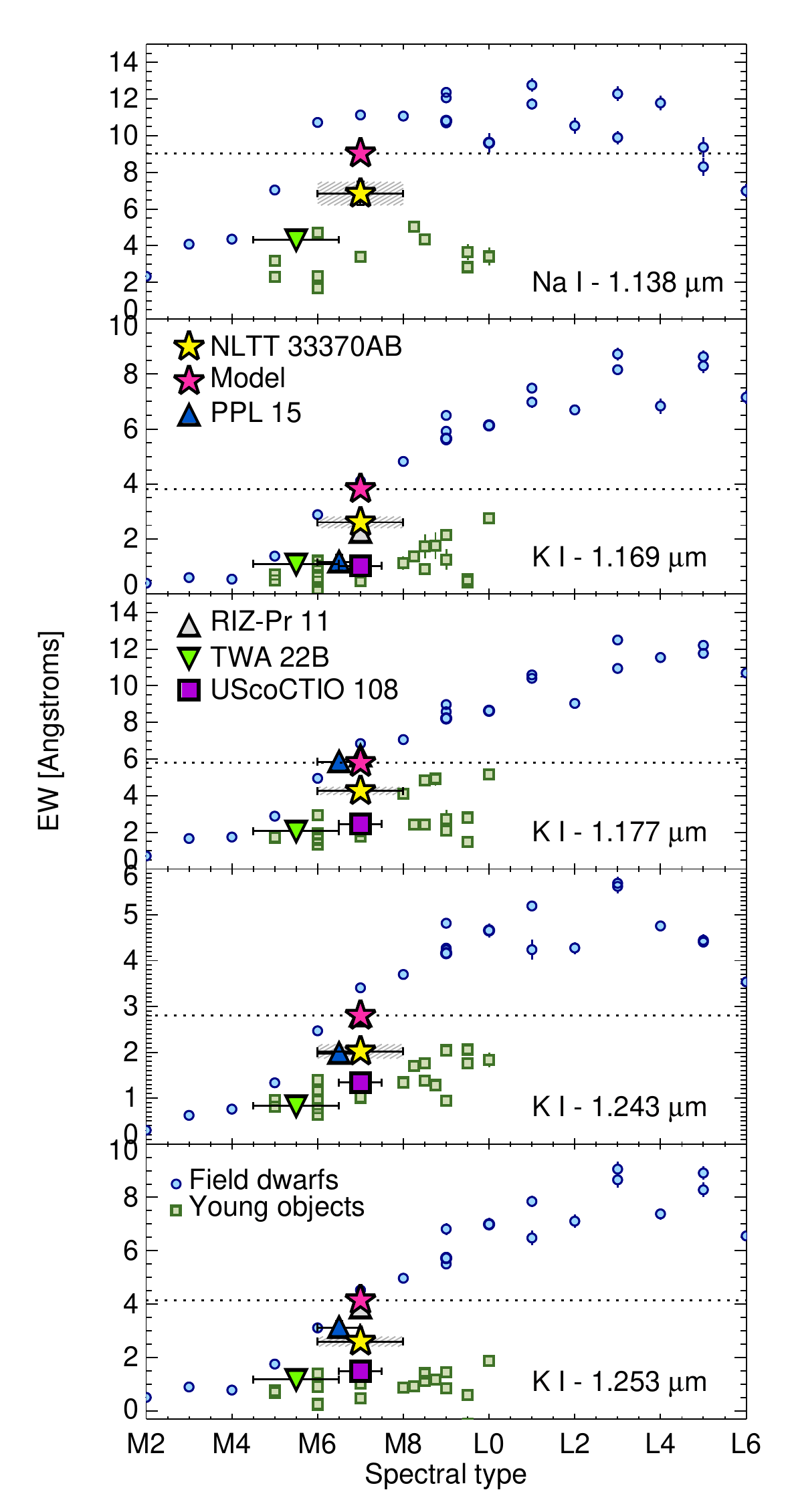}
\caption{Equivalent widths (EWs) of $J$-band Na I and K I lines in young to field age VLM dwarfs.  NLTT 33370AB is designated with a yellow star. Other symbol designations are given in the figure legend and described in the text. The magenta star labeled ``Model" represents the simulated binary shown in Figure~\ref{figjspec} and Figure~\ref{fighkspec}.  The EW of each alkali line in the spectrum of NLTT 33370AB is consistent with an age older than the young dwarf sequence and younger than the field.  The EW measurements reinforce our visual interpretation that the binary's peculiar spectral features are indicative of reduced surface gravity and a young age.} \label{figalkEW}
\end{figure}

{Our comparisons to similar spectral type standards to estimate the age from gravity sensitive features provide meaningful constraints but are qualitative.  Here we take a more quantitive approach and use measured EWs of alkali lines to investigate the binary's age. In Figure~\ref{figalkEW} we show the EWs of Na I and K I lines in the $J$-band spectrum of the binary compared to a selection of the standards plotted in Fig.~\ref{figjspec} and samples of young ($\lesssim$50 Myr) and field VLM dwarfs.  The young dwarf spectra are drawn from the samples presented in \citet{2013arXiv1306.3709B}, \citet{Rice10}, \citet{2008MNRAS.383.1385L}, and  \citet{Slesnick04}.  The spectral types of the \citet{2008MNRAS.383.1385L} objects were reassessed in \citet{2013arXiv1306.3709B}.  The field VLM dwarfs were drawn from \citet{Cushing05} and \citet{2003ApJ...596..561M}.  All spectra used to measure the EWs were reduced to a resolution of 1400 and placed on a common wavelength scale.  The method used to compute the EWs is the same as that described in \citet[][]{2013arXiv1306.3709B} and uses the same pseudo-continuum and line integration regions. These regions were chosen to be insensitive to systematic effects in SINFONI spectra and are thus very well suited to our data.

The computed EWs for each line in the spectrum NLTT 33370AB are systematically smaller than those of the field dwarfs.  The widths of these lines are also larger than those of the young U Sco comparison spectrum.  The EWs of the alkali lines in Pleiades member PPL 15 and Praeseppe member RIZ-Pr 11 exhibit significant scatter, likely due to the lower SNR of these spectra (see Fig.~\ref{figjspec}). We also show in Fig.~\ref{figalkEW} how the alkali lines of the simulated binary presented in Fig.~\ref{figjspec} compare to those of NLTT 33370AB. The simulated binary appears older than NLTT 33370AB in each case and most consistent with the field sequence. The analysis of the alkali line EWs in NLTT 33370AB confirms our visual interpretation of the spectral features. 

Comparison of our EW measurements to the gravity classification scheme introduced by \citet{Allers2013} suggests an intermediate gravity for NLTT 33370AB with an associated age $\sim$50-200 Myr.  Although our computation of the line EWs is not identical to the method of \citet{Allers2013}, the large separation of the binary from the field age sequence in Fig.~\ref{figalkEW} indicates that the small errors introduced by differences in the EW calculation would have little effect on this outcome. This classification is consistent with the age constraints inferred from visual comparison of our spectra to standards of known age.

Similar to the checks we describe in Section~\ref{SINFONIanalysis}, \citet{Allers2013b} perform a more general procedure to show that in no case can an unresolved, field age binary conspire to provide EWs indicative of low surface gravity. Although our use of an unresolved spectrum is not ideal, \citet{Allers2013b} also showed that the strengths of gravity sensitive features are relatively unaffected by constructing artificial binaries out of known young dwarfs.  Thus, the peculiar spectral features of NLTT 33370AB are best explained by a reduced surface gravity and hence a young age. Ultimately, resolved spectra of each component will provide stronger constraints on the age of each component.

\subsubsection{Lithium Depletion}

The detection of strong Li in NLTT 33370AB can also be used as an additional age constraint.  Models predict that low-mass stars and the highest-mass brown dwarfs rapidly deplete their initial Li in their interiors, although there are some discrepancies between the models \citep[e.g.][]{Chabrier96, Burrows97, Baraffe98, Yee2010}.  To explore Li depletion derived age constraints for the components of this binary and to probe the systematic effects introduced by choice of model, we first consider the timescales for lithium depletion in stars of similar temperature.

 \citet{Palla2007} provide model curves of growth for the Li 6708~\AA~line in young stars with temperatures ~3000-4000 K.  Their models predict that the Li EW of a young star with a temperature of ~3000K will only be reduced by $\sim$100 m\AA~as Li is depleted by a factor of 100 (100$\times$) from the primordial abundance.  Even at 1000$\times$ primordial depletion, the line is predicted to have an EW$\approx$150 m\AA.  EWs this small should still be detectable in high-resolution, high SNR spectra. Palla et al.'s curves of growth do not predict the EW of the line after 1000$\times$ primordial depletion, but their may be a considerable amount of time before the Li is depleted beyond the limits of spectroscopic detection. \citet{Yee2010} use EW measurements and the curves of growth from \citet{Palla2007} to estimate the Li depletion fractions of late-type stars in the $\sim$10 Myr $\beta$ Pictoris moving group.  Their empirical results verify that the Li line is detectable with EWs as small as $\sim$100 m\AA.  The Yee \& Jensen sample also contains some stars where Li is not detected with upper EW limits of 10-30 m\AA. They extrapolate depletion fractions for these stars $\gtrsim$10000$\times$ primordial.  Thus, a combination of model predictions and empirical measurements suggests that the Li 6708~\AA~line in cool, young stars should become undetectable once it has been depleted by $\sim$10000$\times$ the primordial abundance.     

Since our high-resolution optical spectrum is blended, the true pseudo-continuum of each component is not accessible and direct estimation of the Li depletion fractions in NLTT 33370 A and B following the methods of \citet{Yee2010} are not possible. We therefore compare the derived component luminosities (see Section \ref{phys_param}) over the age range estimated from gravity sensitive features to Li depletion predictions from \citet{Burrows97} and \citet{Baraffe98}\footnote{Using Phoenix/NextGen model atmospheres; http://phoenix.ens-lyon.fr/Grids/NextGen/}  models.  Following the previous discussion, we linearly interpolate the masses and ages where the Li line is predicted to be depleted by 100$\times$ and 10000$\times$ the primordial abundance in each of the models.  The top panel of Figure~\ref{figlidep} shows the predictions of the Burrows et al. models.  The solid red line represents the model predictions for Li to be depleted from its primordial value by 100$\times$.  The dashed red line is the prediction for 10000$\times$ depletion.  During the time between these depletion fractions, $\sim$10-15 Myr, Li is rapidly depleted beyond detectability. Since Li is detected in the system, the Burrows et al.~models predict that NLTT 33370 A and B are $\lesssim$115 Myr and have masses $\mathrm{\lesssim95~M_{Jup}}$.

The bottom panel of Fig~\ref{figlidep} shows Li depletion predictions from the Baraffe et al.~models.  Here, The green lines represent 100$\times$ and 10000$\times$ primordial Li depletion.  The time scale between these two depletion fractions is longer compared to the Burrows et al.~models, $\sim$30-50 Myr. The models predict that the binary components are $\lesssim$150 Myrs old and have masses $\mathrm{\lesssim100~M_{Jup}}$. The two models thus predict the same masses for the components but upper age limits that are discrepant by about 30 Myrs.  However, since we measure an $\sim$500 m\AA~Li EW from our unresolved spectrum, it is likely that at least one of binary components has depleted its primordial Li by $<<$10000$\times$ and the ages inferred from model depletion timescales are true upper limits.  Never the less, these age constraints from Li depletion are similar to those derived from gravity diagnostics and strengthen the case of a young age for NLTT 33370 A and B.  Also, \citet{Stauffer98} presented empirical evidence that Pleiades VLM dwarfs with types later than M6.5 all exhibit Li.  The lack of resolved spectra prevent individual spectral type estimates for NLTT 33370 A and B, but the combined optical type of M7.0$\pm$0.5 and observed near-IR contrasts indicate that the more massive component cannot be of much earlier spectral type.  Thus, empirically, the detection of Li also suggests an age comparable to the Pleiades.} 

\subsubsection{Hertzsprung-Russell Diagram}

Although our ultimate aim is to use NLTT 33370AB as benchmark to test stellar evolution models at low-masses and young ages, evolutionary model comparisons are useful as a consistency check for the generally young age inferred from empirical data. We first compare the positions of NLTT 33370 A and B to model predictions in a Hertzsprung-Russell (HR) diagram.  Since our near-IR photometry indicates that the components are nearly equal mass, we assume in this study their spectral types are approximately equal with a one subclass uncertainty (M7$\pm$1) and use two empirical relations to estimate their effective temperatures. The main-sequence calibrated temperature scale of \citet{Pecaut2013} provides an effective temperature of $\sim$2700$\pm$100 K for the components.  The scale of \citet{2003ApJ...593.1093L}, which is calibrated for very young ($\sim$few Myr) low-mass dwarfs, provides effective temperatures of $\sim$2900$\pm$100 K.  We compared the component luminosities in these temperature ranges to evolutionary grids from \citet{Baraffe98} models. Figure~\ref{figHRD} shows the model comparison using the temperatures estimated from the Luhman et al. scale. The models predict ages $\sim$30-150 Myr and masses $\mathrm{60-100~M_{Jup}}$.  The temperatures derived using the \citet{Pecaut2013} scale predict ages of $<$50 Myr and masses $\mathrm{<60~M_{Jup}}$.  Our previous analyses rule out the majority of this younger age range and we prefer the temperature estimates from the Luhman et al. scale. Thus, the model HR diagram places similar constraints on the age and mass of the system compared to other age dating methods. 

\subsubsection{Main Sequence Contraction}

Multiple comparisons of gravity sensitive atomic, molecular, and continuum features to empirical spectra of late-M dwarfs across a range of ages provide strong evidence that NLTT 33370 A and B have reduced surface gravities and have not yet reached their final contraction radii. We can combine this information with the component luminosities to place additional model based constraints on the age of the system. For the component luminosities derived in Sec~\ref{phys_param}, we choose as an initial guess ages between 600 and 100 Myr as input to the \citet{Baraffe98} models. Since the luminosities remain constant, guessing different system ages leads to different component masses and different contraction timescales. Given the uncertainties in our luminosities and systematic errors in the models, we define the point where the radius reaches 90\% of the radius at 10 Gyr as the final contraction radius.

Following this method, if the components are 600 Myrs old, the models predict they should have reached their final contraction radii more than 300 Myrs ago. For an age of 400 Myr, the components should have reached their final radii $\sim$100 Myrs ago. If the components are 200 Myrs old, they should have reached their final radii $\sim$50 Myrs ago. At an age of 100 Myr, NLTT 33370 A and B have not yet reached their final radii and still have extended atmospheres. Therefore, combined with the spectroscopic evidence for low surface gravity, this analysis places a model dependent upper limit on the system age of $\sim$200 Myr.

\subsubsection{Age Summary}

Our visual comparisons of NLTT 33370AB's optical and near-IR spectra to dwarfs of known age reveal that gravity sensitive alkali lines, molecular bands, and spectral morphology are consistent with an age younger than the field.  The detection of Li 6708~\AA~absorption in the binary's optical spectrum and comparisons to observed data and model depletion timescales also suggest a young age.  As a consistency check, we compare the HR diagram position of the components to models and consider predicted main sequence contraction timescales. Each of these analyses provide the following constraints on the age of NLTT 33370AB:
\ \\
\begin{itemize}
\item{Gravity diagnostics: $\sim$30 - 300 Myr; alkali line strengths in the $J$-band and the overall morphology of the $H$+$K_s$-bands are most consistent with Pleiades member PPL 15 and AB Doradus group member AB DorC; with ages $\sim$130$\pm$20 Myr.}
\item{Lithium 6708~\AA~absorption: $\lesssim$150 Myr; Li line strength is comparable to that measured in similar spectral type stars with ages $\sim$20-130 Myr. Detection of Li is empirically consistent with Pleiades age for component spectral types of M7$\pm$1. }
\item{HR diagram placement: $\sim$30 - 150 Myr; consistent with the age ranges inferred from gravity diagnostics and Li.}
\item{Contraction timescales: $\lesssim$200 Myr; models predict that VLM dwarfs with the measured luminosities of the components have reached their final radii and should not exhibit spectroscopic signatures of low surface gravity if they are older than 200 Myr.}
\end{itemize}

\noindent{These age ranges derived using different methods illustrate the difficulty in precise age dating of young, VLM dwarfs.  Observational constraints, a lack of reference objects, and complex atmospheres prevent the precision obtained with age dating methods appropriate for earlier-type stars. While our data provide direct evidence for an age most comparable to the Pleiades, unresolved spectra and known  issues with evolution models represent extra sources of age uncertainty.  In light of these complications, and considering the ranges presented above, we adopt an estimated age of $\sim$30-200 Myr for NLTT 33370AB.}

\subsection{Effects of Activity on Age Interpretation}

Empirical evidence indicates that fundamental properties of low-mass stars and brown dwarfs (radius, effective temperature) can be affected by magnetic activity \citep[e.g.][and references therein]{Stassun2012}.  From a theoretical standpoint, changes in these observables are related to the reduction of convective efficiency in the presence of magnetic fields \citep{Chabrier07, MacDonald09}.  The magnetic field strength of NLTT 33370AB is estimated to be large from radio observations, in the range 0.5- 8.0 kG \citep{McLean11}.  We therefore used the empirical relations of \citet{Stassun2012} and the average H$\alpha$ emission reported in \citet{McLean11} as a proxy for magnetic activity to estimate the effects on temperature and radius.  We find that under the influence of magnetic activity, the temperature of the NLTT 33370AB system could be reduced by $\sim$5\% and the radius inflated by $\sim$10\%.  

These results raise some concern in the interpretation of the system age.  If the radii of the components are inflated due to magnetic activity, does this significantly contribute to a reduced surface gravity and the peculiarity of the associated spectral features?  \citet{McLean11} use the measured rotation period from radio and optical monitoring ($\approx$3.9 h) and the projected rotation velocity ($v$sin$i$ = 45 $\pm$ 5 km s$^{-1}$) to place a lower limit of 0.13 R$_{\odot}$ on the radius of the primary component (to which they attribute the measured rotation period). This is $\gtrsim$40\% larger than model predictions for a field age M7 dwarf. We recognize that the rotational velocity was measured from an integrated light spectrum of both components.  Thus, blending of individual spectral features may contribute to line broadening and result in an overestimated $v$sin$i$.  However, these simple considerations indicate that the radius of at least one component is likely inflated much more than estimated by magnetic activity and that a young age also contributes to the inferred low surface gravity. Additionally, our use of the component luminosities for model comparisons are negligibly affected by activity induced radius and temperature changes, because the combined effects approximately cancel out \citep{Stassun2012}.  

Since magnetic activity is related to suppressed convection in the stellar interior, it is also possible this mechanism could directly affect the rate of Li depletion. For instance, studies of solar-type members of the Pleiades revealed a possible correlation between rapid rotation/strong activity and inhibited Li depletion \citep[e.g.][]{King00, Tschaepe01, King10}.  But these results are based on observations of stars with vastly different masses and interior structures than the components of the VLM binary considered here. \citet{Macdonald2010} use models that include the effects of magnetic fields on convective efficiency to explore HR diagram positions and Li depletion timescales in low mass stars.  They find that for temperatures $\lesssim$3000 K magnetic fields have a relatively small influence since the degree of ionization is low. This does not allow for effective coupling of the fields to the plasma and convection remains relatively efficient.  Although we cannot rule out that the combined effects of rapid rotation and magnetic activity may contribute to inhibited Li depletion in NLTT 33370AB, we interpret the similarity between the age constraints from gravity diagnostics and Li as evidence that any such effects are small in this particular case.
    
\subsection{System Physical Parameters}
\label{phys_param}

The near-IR colors of each component are compatible with those of mid-M dwarfs and other young (8-100 Myr), benchmark M5-M9.5 dwarfs (see Figure \ref{figCCd}). However, these comparisons do not allow us to further constrain the component spectral types and system age estimate. To further investigate their properties, we constructed the 1.2-4.7 $\mu$m spectral energy distribution (SED) of NLTT 33370 A \& B from the $JHKL'$-band photometry of the components.  We then compared the SEDs to fluxes predicted by BT-Settl2010 and 2012 atmospheric models for different surface gravities and effective temperatures (see Figure~\ref{figSED}). The fitting procedures and models are described in \cite{2013arXiv1302.1160B}. We chose to use the 2010 release of the models, since they better reproduce the the M-L transition at young ages \citep{2013arXiv1306.3709B}. The best fit atmospheric parameters appear in Table~\ref{atmoparPHOENIX}.  Figure~\ref{figchi2maps} shows the corresponding $\chi^{2}$ maps. The maps indicate that both models give poor constraints on the temperature, and basically no constraint on the surface gravity (95\% confidence level). Nevertheless, the choice of atmospheric models has little impact on the best fit temperatures (100 K scatter). The estimated temperatures are consistent with our previous estimates from the spectral type and roughly correspond to those expected for M4-M7 dwarfs using the \citet{2003ApJ...593.1093L} conversion scale. BT-COND evolutionary models\footnote{http://phoenix.ens-lyon.fr/Grids/BT-Cond/ISOCHRONES/} predict contrasts in the optical of $0.6-0.9$ mag for the estimated system age. These are slightly smaller than those measured by \cite{Law06} for these components.

Simultaneously fitting the individual $z'$ magnitudes measured by \cite{Law06} yields $\mathrm{T_{eff}}=3300$ K for NLTT 33370 A and 2600 K for NLTT 33370 B. The model radius inferred for NLTT 33370 B ($\mathrm{2\:R_{Jup}}$) from this temperature deviates from predictions for the luminosity of such an object (assuming an age of 30-150 Myr). Also, evolutionary models predict higher contrast ratios than observed in the $JHKL'$- bands for a $\sim$700K temperature difference between the components. This aspect of our SED analysis may be explained by a bias toward inflated contrast ratios for tight, $\sim$equal brightness binaries in Lucky Imaging data.  This trend was reported by \citet{Janson2012} in their Lucky Imaging survey of M dwarfs.

We used the $J$-band photometry, bolometric corrections reported in \cite{Liu10} for spectral types of M$7 \pm1$, and the measured distance to the binary to derive $\mathrm{log_{10}(L/L_{\odot})} =  -2.64 \- \pm \- 0.06$ and  $\mathrm{log_{10}(L/L_{\odot}) \- = \- -2.68 \- \pm \- 0.06}$ for NLTT 33370 A and B, respectively.  These luminosities correspond to masses of $113\pm8$  and  $106\pm7\:\mathrm{M_{Jup}}$, respectively, using \cite{Baraffe03} evolutionary models and the age range determined in the previous sections.  These masses are consistent with the upper limit inferred from comparison of Li depletion to the \citet{Burrows97} and \citet{Baraffe98} models. The adopted values appear in Table~\ref{tab1}. 

\subsection{Galactic Kinematics}\label{kinematics}

Many young stars near the Sun are members of kinematic groups, coeval associations of stars with common Galactic kinematics \citep[see reviews by][]{Zuck04_2, Tor08}.  Since precise age determination of an isolated VLM dwarf is difficult, strong kinematic evidence for membership in one of these groups combined with independent constraints for a young age can provide necessary age calibration references. This approach was used by \citet{Allers2013} to assess the age ranges related to their surface gravity classifications. We therefore use the measured kinematics of NLTT 33370AB to explore its possible membership in a young, moving group to provide additional constraints on its age.  

We used the estimated $\mathrm{RV_{sys}}$ and previously measured proper motion and parallax of the binary to calculate the system's Galactic velocities and distances (UVWXYZ) using the procedures described by \citet{Johnson1987}\footnote{We adopt a coordinate system where U/X are positive toward the Galactic center, V/Y are positive in the direction of the Sun's orbit, and W/Z are positive toward the north Galactic pole.}.  We find (UVW) = (-9.6$\pm$0.7, -21.0$\pm$1.2, -11.8$\pm$1.0) km s$^{-1}$ and (XYZ) = (3.4$\pm$0.1, -2.4$\pm$0.3, 15.8$\pm$0.6) pc.  

These values were compared to the distributions of known young ($\lesssim$100 Myr) kinematic groups listed in \citet{Tor08} and \citet{Malo2013}.  The full Galactic kinematics of NLTT 33370AB are similar only to proposed members of the AB Doradus moving group \citep[(UVW)$_{ABD}$ = (-7.1$\pm$1.4, -27.3$\pm$1.3, -13.8$\pm$2.2) km s$^{-1}$, $\sim$100 Myr,][]{Malo2013, Baren13}. NLTT 33370AB is consistent with the group's Galactic distances, but is a 5$\sigma$ outlier in both the (UV) and (VW) projections of Galactic velocity.  The space velocities of the system are also broadly consistent with the $\sim$5 Myr old $\epsilon$ Cha group \citep{Tor08, Murphy2013}, but the estimated age and measured distance rule it out as a possible member.  Further comparison to less well defined moving groups reveals that the kinematics of NLTT 33370AB are most consistent with the proposed members of the Hercules-Lyra association \cite[260$\pm$50 Myrs]{Eisenbeiss2013}.  This putative kinematic group however numbers only seven members meeting all membership criteria and exhibits large dispersions in (UVW) velocity ($>$3.5 km s$^{-1}$).

Thus, the currently available kinematic data does not allow us to suggest NLTT 33370AB is a bona fide member of any well established young kinematic group. We can therefore place no further constraints on its age and conclude only that its kinematics are broadly consistent with other young stars in the Solar neighborhood.  Definitive membership evaluation will require better constraints on its systemic RV.    

\section{Conclusions}

We confirm the binary nature of NLTT 33370 via VLT-NaCo and LBTI-LMIRCam AO imaging with a projected separation of 76$\pm$5 mas (1.24$\pm$0.11 AU).  Multi-epoch imaging reveals a PA change of $\sim$10$^{\circ}$ per month, consistent with a $\sim$3-7 year orbital period.  Comparison of our astrometry to astrometry from the 2005 discovery epoch provides preliminary evidence for a low inclination, eccentric orbit. Re-analyses of previously published optical spectra and new VLT-SINFONI near-IR spectra reveal both weak, gravity sensitive spectral features and strong lithium absorption. We use these data and comparisons to models to constrain the system age to $\sim$30-200 Myr.  

Spectral template fitting and resolved $JHKL'$-band photometry from our follow-up imaging allowed us to estimate $\mathrm{T_{eff}}=3200\pm500$ K for NLTT 33370A and $\mathrm{T_{eff}}=3100\pm500$ K for NLTT 33370B.  However, the fitting routines were unable to constrain their surface gravities.  Model comparisons to component luminosities and the estimated system age suggest masses of 113$\pm$8 and 106$\pm$7  $\mathrm{M_{Jup}}$ for the A and B components respectively.  These preliminary, model based, mass estimates place the components near the substellar boundary.  

We used a cross-correlation analysis of KPNO-Phoenix spectra to estimate the systemic RV of the system and then calculate its Galactic kinematics.  NLTT 33370AB has space velocities consistent with other young stars in the Solar neighborhood, but further follow-up is required to fully assess potential kinematic group membership. Thus, no further constraints on the age of the system could be made.       

NLTT 33370AB is a rare example of a young, VLM binary where the full orbital parameters can be rapidly determined.  The system has many similarities to TWA 22AB in component separation, likely orbital period, and total mass \citep[see][]{Bonnefoy09}.  On the other hand, the NLTT 33370AB is likely much older and provides a new benchmark for calibration of low-mass evolution models.    

 \acknowledgments
 
We thank the anonymous referee for a thorough and constructive review that improved this manuscript. We thank the ESO Paranal staff for conducting the service mode observations presented here. We also thank the LBTI/LMIRCam instrument team for providing support during those observations. The authors are grateful for the support of the KPNO staff, particularly Dick Joyce, Dave Summers, and Kristin Reetz.  We are grateful to Chad Bender for making available his software for the two dimensional cross-correlation of stellar spectra. J.E.S. thanks Roland Gredel, Niall Deacon, Eddie Schlafly, and Taisiya Kopytova for helpful conversations regarding the observations and analyses presented here. The research of Thomas Henning on moving groups and the low-mass stellar population in the solar neighborhood is partly supported by the German Science Foundation SFB 881``The Milky Way System". E.~B.~acknowledges support for this work from the National Science Foundation through Grant AST-1008361. The research leading to these results has received funding from the French ``Agence Nationale de la Recherche" through project grant ANR10-BLANC0504-01, the ``Programme Nationale de Physique Stellaire" (PNPS) of CNRS (INSU). This paper includes data gathered with the 6.5 meter Magellan Telescopes located at Las Campanas Observatory, Chile. This research has made use of the SIMBAD database, operated at CDS, Strasbourg, France. This publication makes use of data products from the Wide-field Infrared Survey Explorer, which is a joint project of the University of California, Los Angeles, and the Jet Propulsion Laboratory/California Institute of Technology, funded by the National Aeronautics and Space Administration.  Based on observations made with the NASA/ESA Hubble Space Telescope, and obtained from the Hubble Legacy Archive, which is a collaboration between the Space Telescope Science Institute (STScI/NASA), the Space Telescope European Coordinating Facility (ST-ECF/ESA) and the Canadian Astronomy Data Centre (CADC/NRC/CSA). This research has made use of the Keck Observatory Archive (KOA), which is operated by the W. M. Keck Observatory and the NASA Exoplanet Science Institute (NExScI), under contract with the National Aeronautics and Space Administration.



{\it Facilities:} \facility{VLT:Yepun (NaCo, SINFONI)}, \facility{LBT (LBTI-LMIRCam)}, \facility{Mayall (Phoenix)}, \facility{MDM:McGraw-Hill (MkIII)}, \facility{MDM:Hiltner (MkIII)}, \facility{Magellan:Clay (MIKE)}

\bibliography{refs}




\begin{table*}
\begin{center}
\caption{Summary of NLTT 33370AB properties \label{tab1}}
\begin{tabular}{lllll}
\tableline\tableline
 & NLTT 33370AB & NLTT 33370A & NLTT 33370B & Reference \\
\tableline
$\alpha_{(J2000)}$ (deg.) & 198.584973   & \dots  & \dots  & 2 \\ 
$\delta_{(J2000)}$ (deg.) & +13.333663   &  \dots  & \dots  & 2 \\
$\mu_{\alpha}$ (mas yr$^{-1}$) & $-244\pm8$  & \dots  & \dots & 2 \\
$\mu_{\delta}$ (mas yr$^{-1}$) & $-186\pm8$  & \dots  & \dots  & 2 \\
d (pc) &  $16.39\pm0.75$ & \dots & \dots  &  1 \\
RV (km s$^{-1}$) & -$10.4\pm1.0$   & -20.2$\pm$1.2 & -0.5$\pm$1.5   & 2 \\
$z'$(mag) &  $11.51$  &  $11.90\pm0.32$  &  $12.8\pm0.21$   &  2, 3  \\
$J$ (mag) &  $9.754\pm0.022$  &  $10.46\pm0.11$  &  $10.56\pm0.11$   &  2, 4  \\
$H$ (mag) & $9.175\pm0.032$  &  $9.91\pm0.10$ & $9.94\pm0.10$  & 2, 4 \\
$K_{s}$ (mag) &  $8.794\pm0.018$  &  $9.50\pm0.21$   & $9.60\pm0.21$   &  2, 4 \\
$L'$ (mag) & $\sim$8.56  & $9.30\pm0.20$  & $9.30\pm0.20$ & 2, 5 \\
Age (Myr) & $\sim$30-200 Myr & \dots  & \dots   & 2 \\ 
$\mathrm{log(L/L_{\odot})}$ (dex) &   $-2.36\pm0.09$ &  $-2.64\pm0.06$ &  $-2.68\pm0.06$    &  2 \\
$\mathrm{T_{eff}}$ (K)   &  \dots    &   $3200\pm500$  & $3100\pm500$  & 2 \\
Mass ($\mathrm{M_{Jup}}$)  &  $219\pm15$  & $113\pm8$  &  $106\pm7$   &    2  \\
\tableline
\end{tabular}
\tablecomments{1 - L\'epine et al. 2009; 2 - This work; 3 - Law et al. 2006 ; 4 - Cutri et al. 2003 ; 5 - Wright et al. 2010, Cutri et al. 2012}
\end{center}
\end{table*}

\begin{table*}[!h]
\caption{Astrometry for NLTT 33370AB}
\label{Table:astrometry}
\centering
\begin{tabular}{llllll}
\hline
UT Date    & True North & Platescale &     $\rho$ &   PA  \\
(DD/MM/YY)	&  (deg)  &	(mas/pixels) 		 &		(mas)    &  (deg)       \\
\hline \hline 
10/02/13  & $-0.59\pm0.29$  &  $13.21\pm0.11$  &   $76\pm5$   &   $204.9\pm0.3^{a}$   \\
13/03/13 &  $-0.61\pm0.25$  &  $13.29\pm0.06$ &  $76\pm6$  &  $215.2\pm1.0^{a}$  \\
19/04/13    & $-0.31\pm0.20$ & $10.74\pm0.05$ &   $73\pm4$   &   $226.7\pm1.6^{a}$      \\
\hline
\end{tabular}
\tablecomments{a - Due to uncertainty in component assignments, may be 180$^{\circ}$ out of phase.}
\end{table*}

\begin{table*}
\caption{Best-fit atmospheric parameters for NLTT 33370 A \& B}
\label{atmoparPHOENIX}
\centering
\begin{tabular}{l|llll|lllll}
\hline \hline 
Atmospheric model			&    $\mathrm{Teff_{A}}$		&	 $\mathrm{log\:g _{A}}$	& $\mathrm{R_{A}}$		& $\mathrm{\chi^{2}_{A}}$&    $\mathrm{Teff_{B}}$		&	 $\mathrm{log\:g _{B}}$	& $\mathrm{R_{B}}$		& $\mathrm{\chi^{2}_{B}}$	\\   			
													&  (K)					&		$\mathrm{(cm.s^{-2}}$)	&	($\mathrm{R_{Jup}}$)		& &   (K)					&		$\mathrm{(cm.s^{-2}}$)	&	($\mathrm{R_{Jup}}$)	\\	
\hline
BT-Settl 2010					&  3200						&	5.0			&	1.55		&	0.83	&  3100						&	4.0			&	1.58		&	0.20	\\
BT-Settl 2012			 & 	3100						&	4.5			&	1.62		&	0.69	& 	3100						&	4.0			&	1.57		&	0.21	\\
\hline
\end{tabular}
\end{table*}
\ \\
\ \\


\begin{figure}[!htb]
\epsscale{0.6}
\plotone{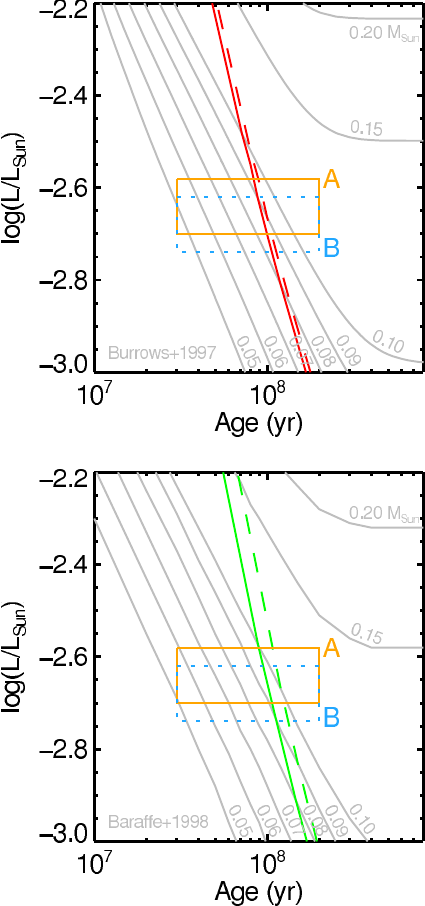}
\caption{Top: Comparison of NLTT 33370 A and B to \citet{Burrows97} models. The solid orange and dotted blue rectangles show the estimated ranges of luminosity and age (from gravity sensitive diagnostics) for the NLTT 33370 A and B respectively.  The gray lines show a series of models with masses in units of M$_{\odot}$.  The thick red lines represent the boundary where Li is depleted by 100X (solid red) and 10000X (dashed red) the primordial value. Observations described in the text suggest that during the $\sim$10-15 Myr between these two lines, the Li absorption feature becomes undetectable in high-resolution spectra. The \citet{Burrows97} models predict an age upper limit of $\sim$120 Myr for the components.  Bottom: Same comparison to predictions of Li depletion from \citet{Baraffe98} models.  The predicted time scale to reach 10000X primordial Li depletion (dashed green) is longer in the Baraffe et al. models, $\sim$30-50 Myr.  The \citet{Baraffe98} models therefore place a larger upper limit of $\lesssim$150 Myr on the age of the components.  Since Li is detected in the system, the models provide an overall age constraint $\lesssim$150 Myr and masses of $\lesssim$0.1 M$_{\odot}$. This age estimate is consistent with the range derived from gravity diagnostics.} \label{figlidep}
\end{figure}

\begin{figure}[!htb]
\epsscale{1.0}
\plotone{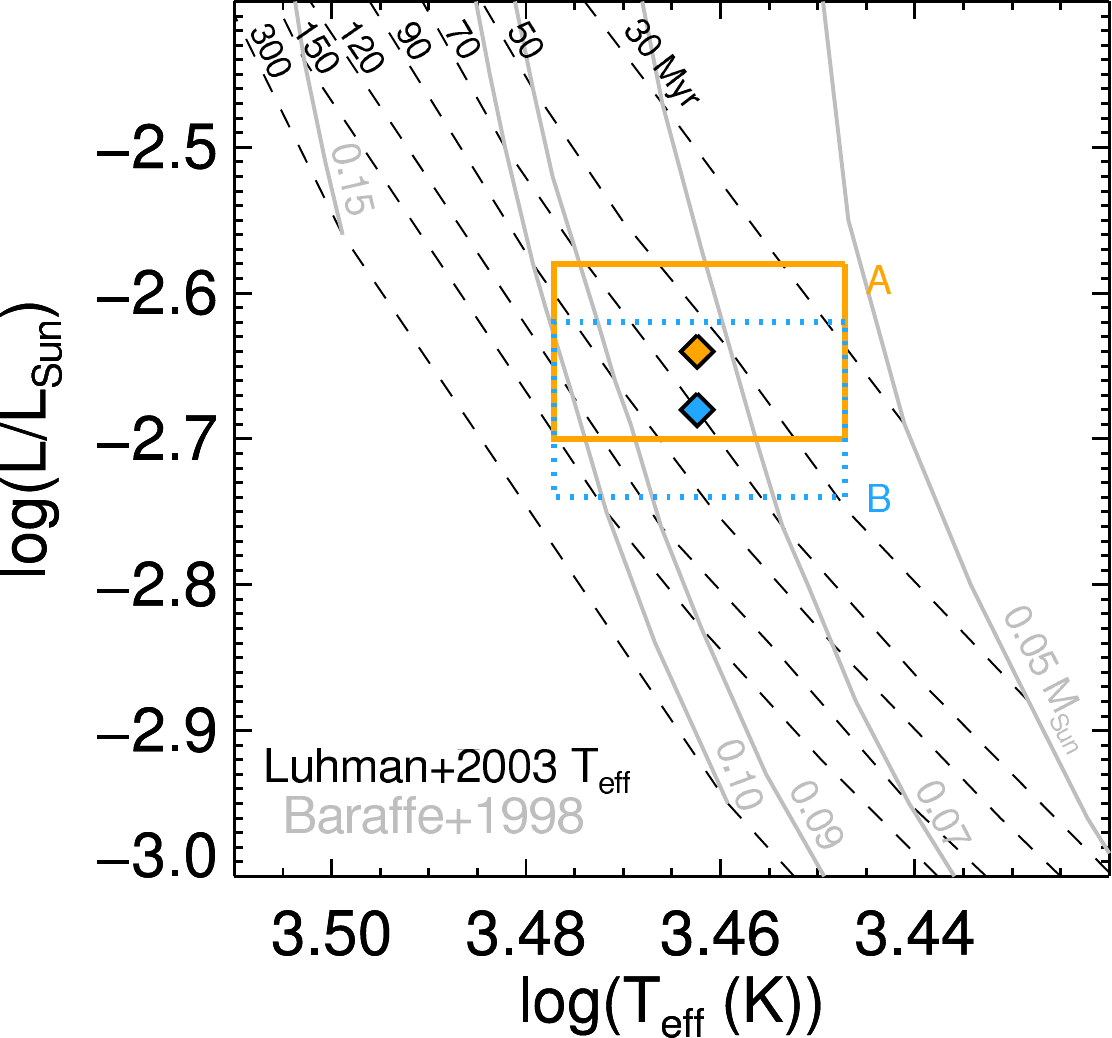}
\caption{Hertzsprung-Russel diagram position of NLTT 33370 A (orange) and B (blue) compared to predictions from \citet{Baraffe98} models.  The gray lines show models of constant mass in units of M$_{\odot}$. The dashed black lines are isochrones.  The estimated component luminosities are based on our resolved $J$-band photometry and the bolometric corrections described in the text.  To estimate the effective temperatures of the components, we use a conservative spectral type of M7$\pm$1 for each and the temperature conversion scale of \citet{2003ApJ...593.1093L}.  When combined with the luminosity estimates,  the range of temperatures estimated for each component are consistent with ages $\sim$30-150 Myr and masses that straddle the hydrogen burning limit.  These estimates are consistent with those from other empirical and model constraints.} \label{figHRD}
\end{figure}

\begin{figure}[!h]
\epsscale{1.0}
\plotone{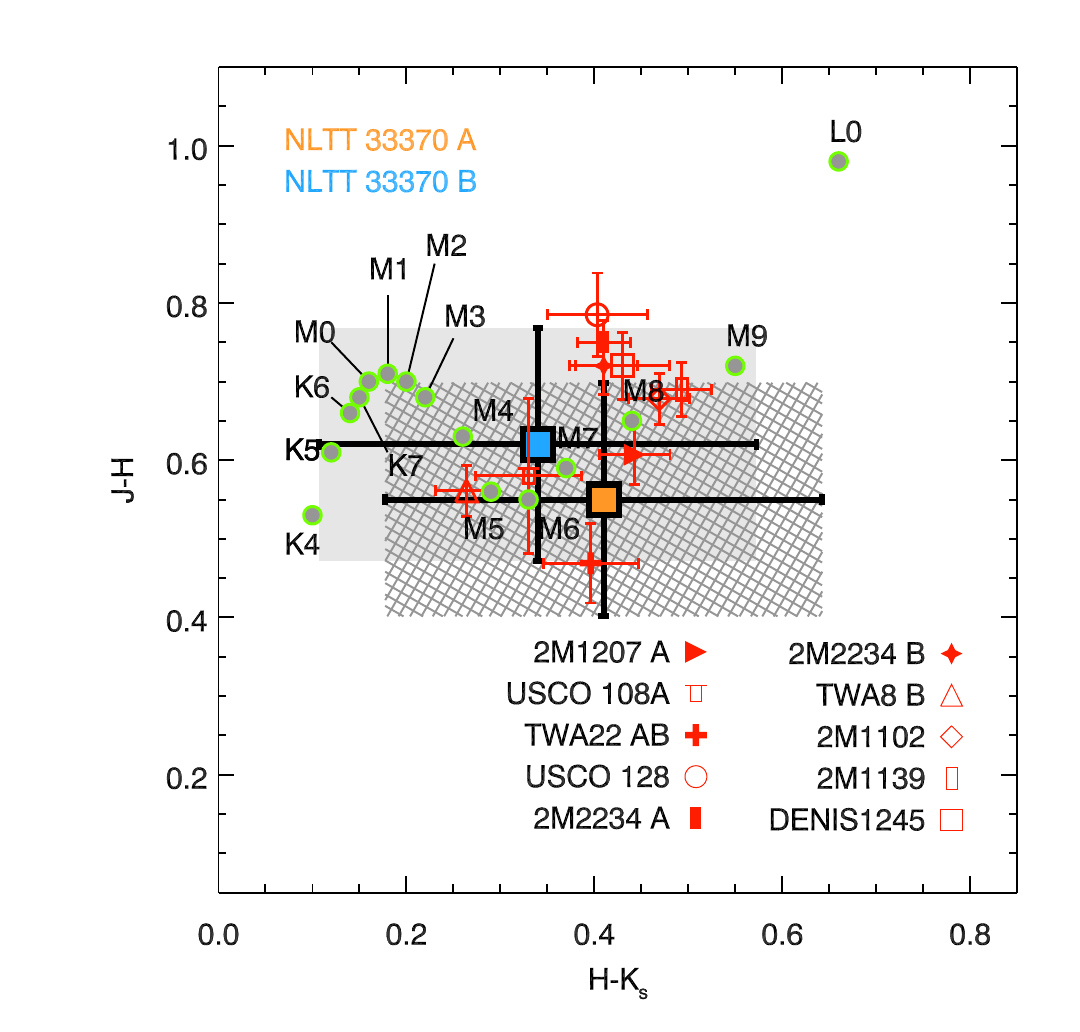}
\caption{Comparison of NLTT 33370 A and B near-IR colors to those of typical young K4-L0 dwarfs \citep{2010ApJS..189..353L} and of young (8-100 Myr) M5-M9.5 companions, primaries, or binaries with accurate photometry. \label{figCCd}}
\end{figure}

\begin{figure}[!h]
\epsscale{1.0}
\plotone{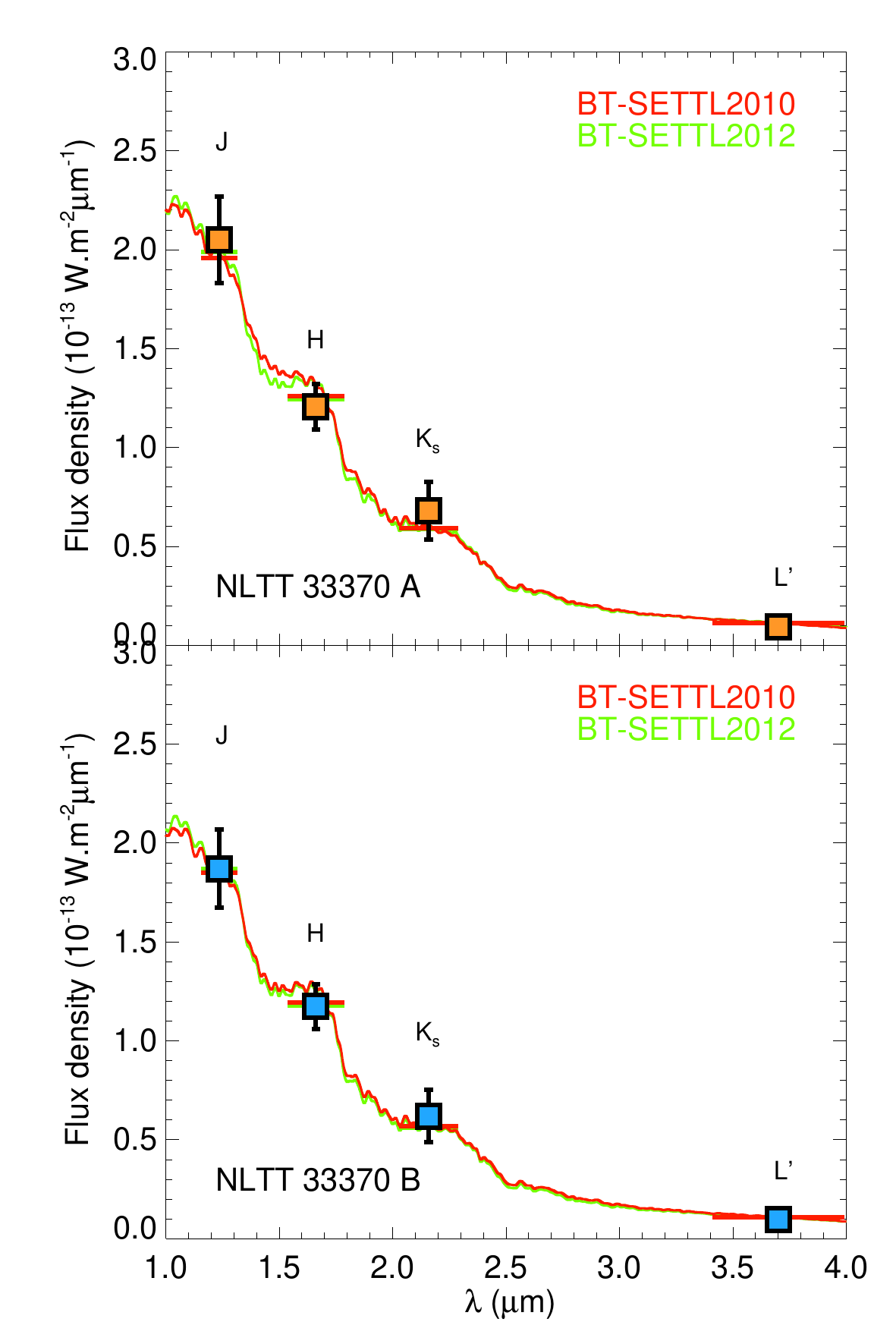}
\caption{Comparison of NLTT 33370 A and B spectral energy distributions to best fit synthetic fluxes (laying bars) generated from the 2010 and 2012 releases of the BT-Settl atmospheric models. The corresponding model spectra are overlaid for comparison. \label{figSED}}
\end{figure}

\begin{figure*}[!htb]
\centering
\begin{tabular}{cc}
\includegraphics[scale=0.28]{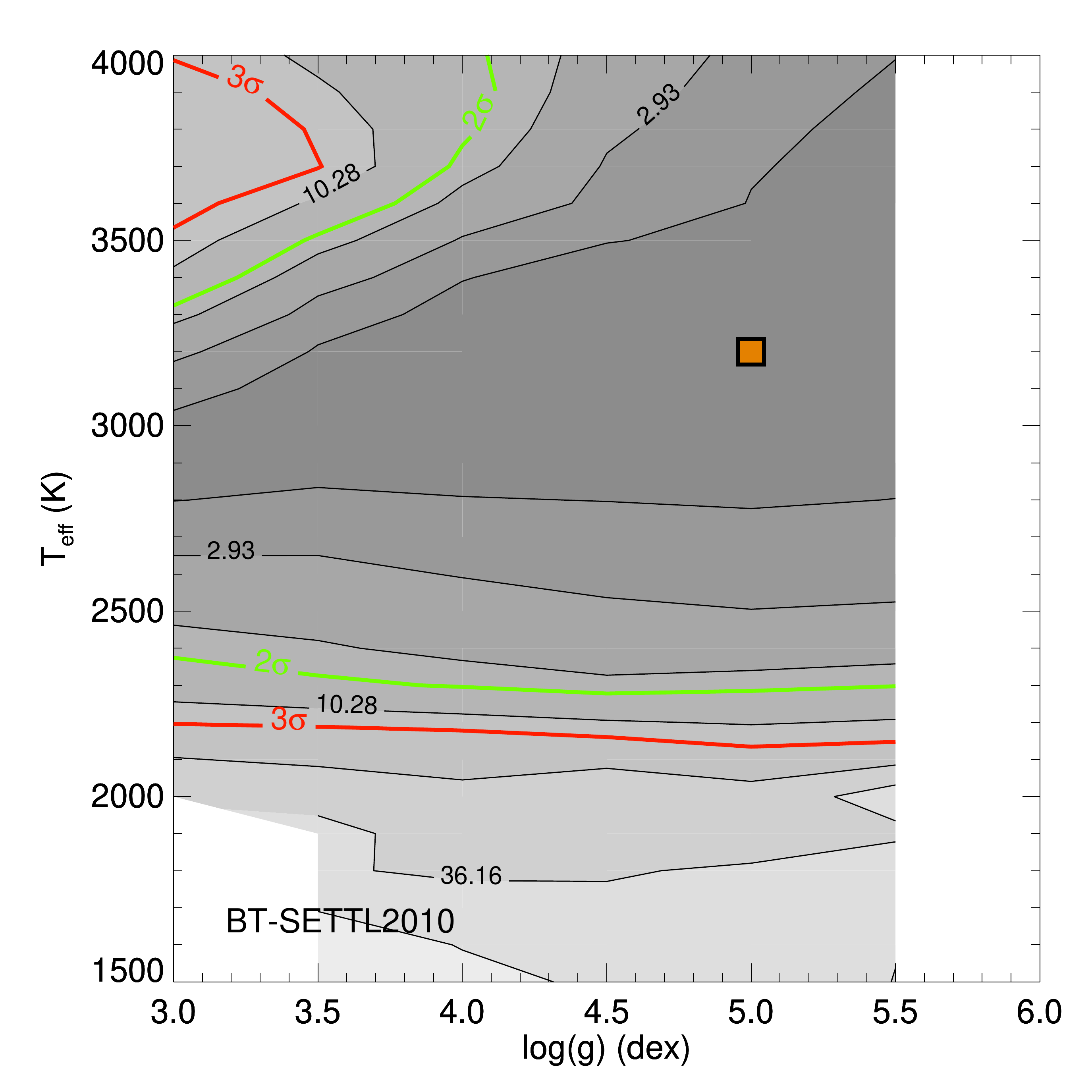} &
\includegraphics[scale=0.28]{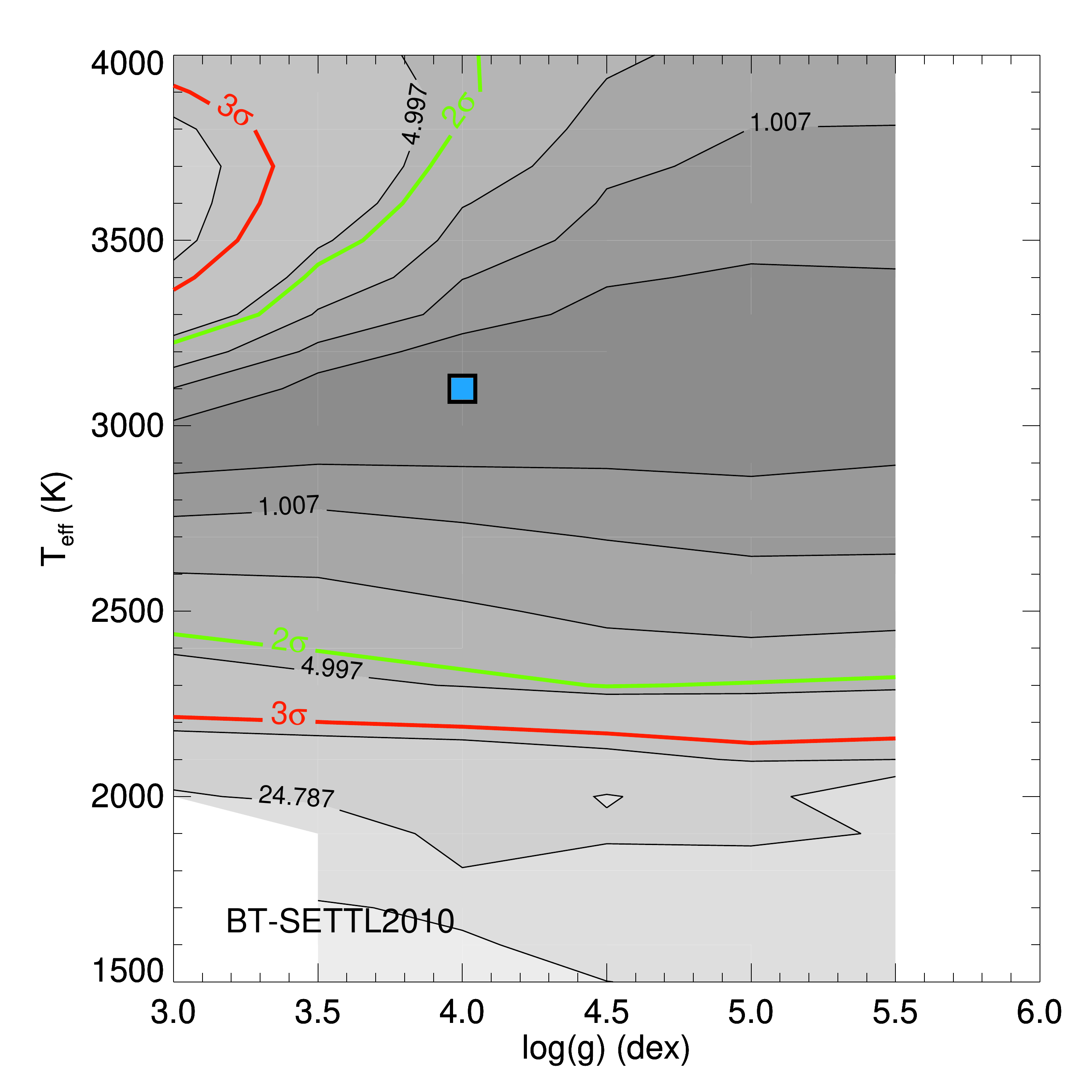} \\
\includegraphics[scale=0.28]{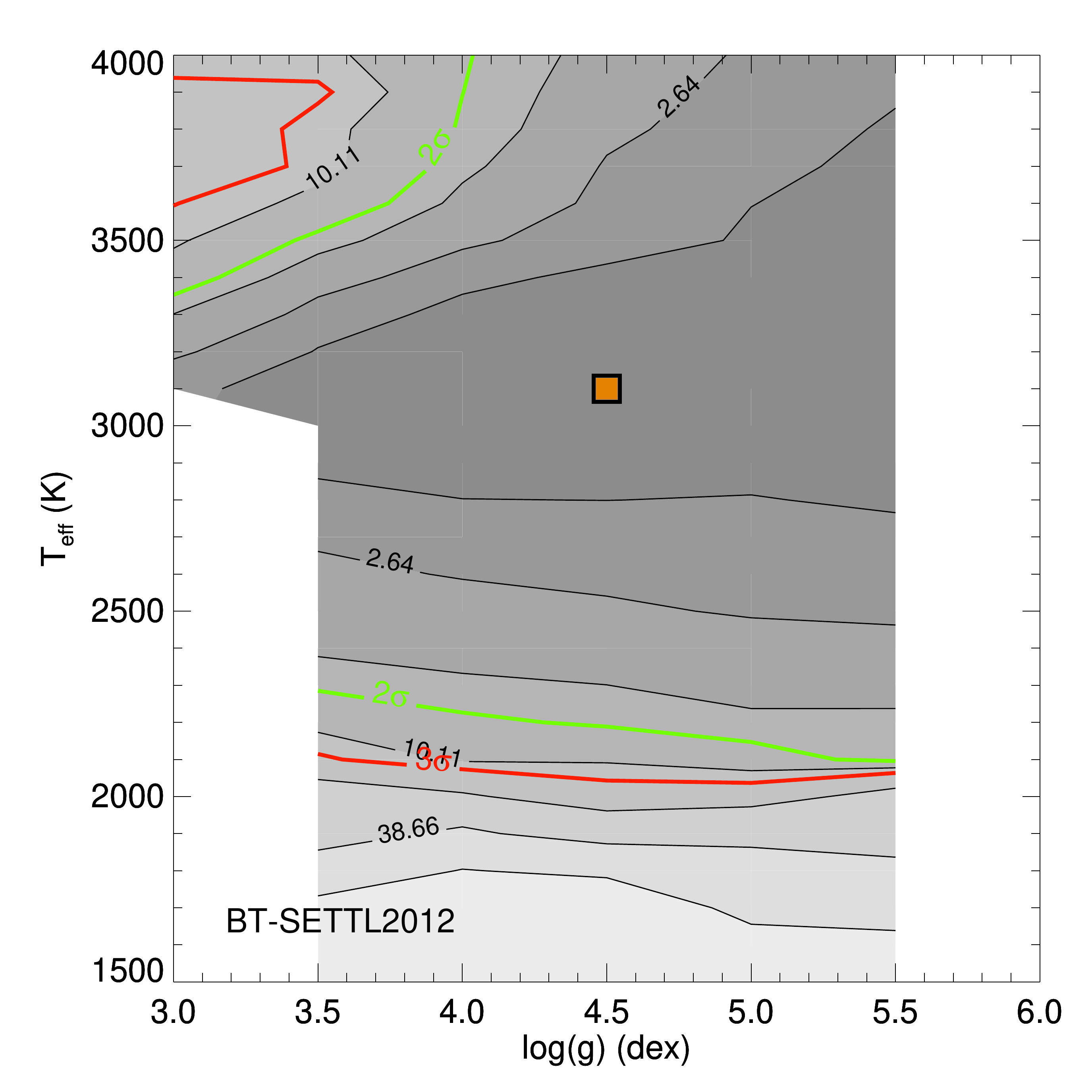} &
\includegraphics[scale=0.28]{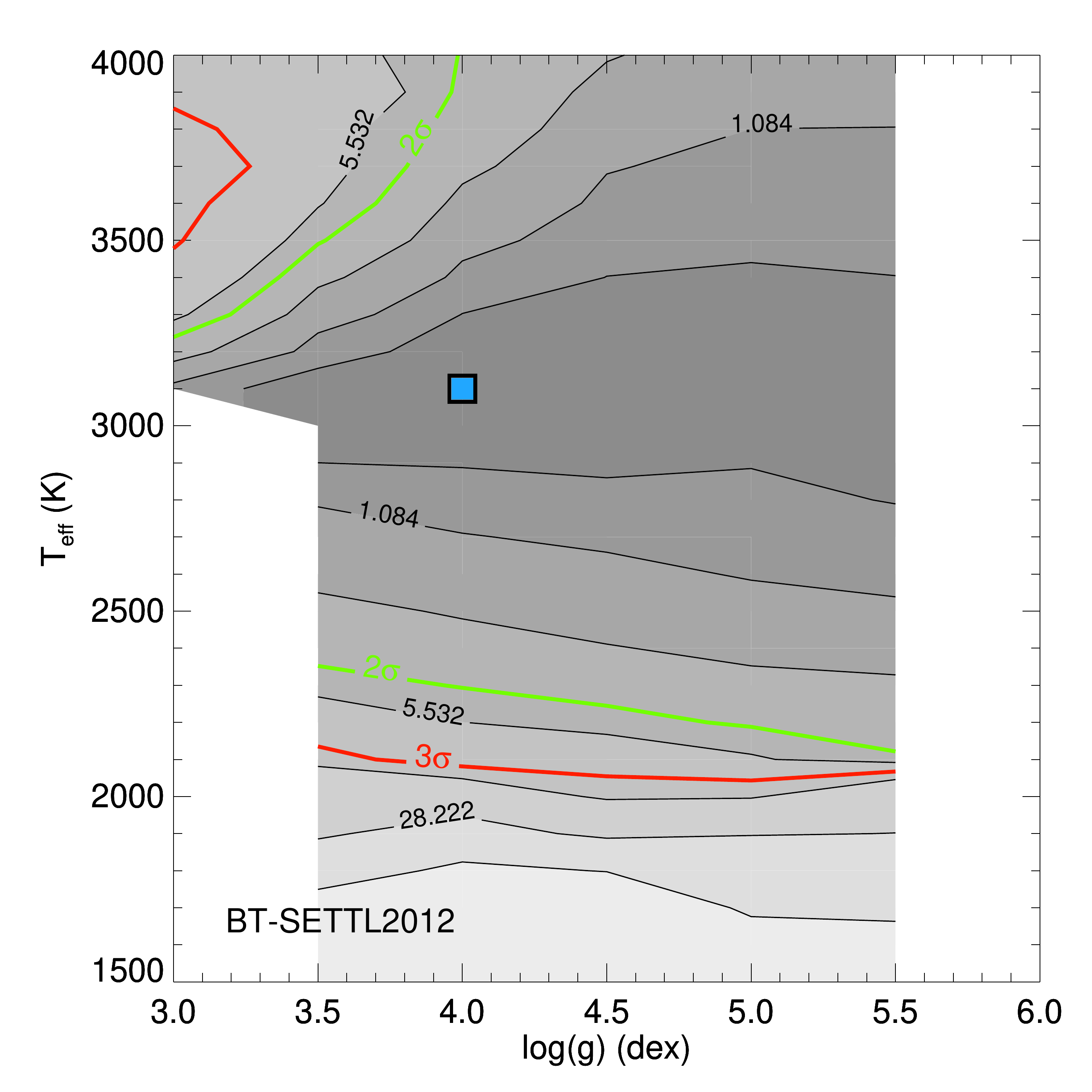} \\
\end{tabular}
\caption{$\chi^{2}$ maps corresponding to the comparison of  the  spectral energy distribution of NLTT 33370 A (left column) and B (right column) to synthetic fluxes derived from BT-Settl 2010 (top row), and BT-Settl 2012 (bottom row) atmospheric models for given log g and $\mathrm{T_{eff}}$.  We overlay contours corresponding to 2$\sigma$ (green) and 3$\sigma$ (red) confidence levels and the location of the minima (squares) in the maps. \label{figchi2maps}}
\end{figure*}







\clearpage


\end{document}